\documentclass{aastex62}

\usepackage{gensymb}

\def\las{\mathrel{\hbox{\rlap{\hbox{\lower3pt\hbox{$\sim$}}}\hbox{\raise2pt\hbox{$<$}}}}}
\def\gas{\mathrel{\hbox{\rlap{\hbox{\lower3pt\hbox{$\sim$}}}\hbox{\raise2pt\hbox{$>$}}}}}

\usepackage[colorinlistoftodos]{todonotes}
\let\Oldtodo\todo
\renewcommand{\todo}[1]{\Oldtodo[inline]{#1}}

\graphicspath{{./}{figures/}}

\received{2019 November 6}
\revised{2020 January 13}
\accepted{2020 February 4 by the Astronomical Journal}

\shorttitle{The Dynamics of Interstellar Asteroids and Comets Within the Galaxy}
\shortauthors{Hallatt and Wiegert}

\begin{document}

\title{The Dynamics of Interstellar Asteroids and Comets within the Galaxy: an Assessment of Local Candidate Source Regions for 1I/'Oumuamua and 2I/Borisov}

\correspondingauthor{Tim Hallatt}
\email{thallatt@physics.mcgill.ca}

\author[0000-0003-4992-8427]{Tim Hallatt}
\affil{Department of Physics and Astronomy, The University of Western Ontario, London, Ontario, ON N6G 2V4, Canada}
\affil{Department of Physics, McGill University, Montr\'eal, Qu\'ebec, QC H3A 2T8, Canada}
\affil{McGill Space Institute; Institute for Research on Exoplanets (iREx); Montr\'eal, Qu\'ebec, Canada}
\author[0000-0002-1914-5352]{Paul Wiegert}
\affil{Department of Physics and Astronomy, The University of Western Ontario, London, Ontario, ON N6G 2V4, Canada}
\affiliation{The Institute for Earth and Space Exploration (IESX), London, Ontario, Canada}



\begin{abstract}

The low velocity of interstellar asteroid 1I/'Oumuamua with respect to our galaxy's Local Standard of Rest implies it is young. Adopting the young age hypothesis, we assess possible origin systems for this interstellar asteroid and for 2I/Borisov, though the latter's higher speed means it is unlikely to be young.

First, their past trajectories are modelled under gravitational scattering by galactic components ('disk heating') to assess how far back one can trace them. The stochastic nature of disk heating means that a back-integration can only expect to be accurate to within 15 pc and 2 kms$^{-1}$ at -10 Myr, dropping steeply to 400 pc and 10 kms$^{-1}$ at -100 Myr, sharply limiting our ability to determine a precise origin. Nevertheless, we show `Oumuamua's origin system is likely currently within 1 kpc of Earth, in the local Orion Arm. 

Second, we back-integrate 'Oumuamua's trajectory to assess source regions, emphasizing young systems and moving groups. Though disk heating allows for only a statistical link to source regions, `Oumuamua passed through a considerable subset of the Carina and Columba moving groups when those groups were forming. This makes them perhaps the most plausible source region, if 'Oumuamua was ejected during planet formation or via intra-cluster interactions. 

We find three stars in the Ursa Major group, one brown dwarf, and seven other stars to have plausible encounters with 2I/Borisov, within 2 pc and 30 kms$^{-1}$. These encounters' high relative speeds mean none are likely to be the home of 2I/Borisov.

\end{abstract}



\section{Introduction} \label{sec:intro}

The first detection of an interstellar object (ISO) occurred on October 19, 2017 \citep{meewermic17}. The Panoramic Survey Telescope And Rapid Response System (Pan-STARRS1) performed the initial detection. Dubbed \textit{'Oumuamua}, loosely meaning \textit{Messenger From Afar}, its discovery has ushered in a new era in the study of planetary objects and planet formation. 

The population of known interstellar objects such as 'Oumuamua has been predicted to grow significantly over the next few years with the advent of the Large Synoptic Survey Telescope \citep{morturloe09}. ISOs therefore represent a new class of object, and may provide unprecedented insight into the dynamics and materials of the systems which produce them. By studying their compositions and dynamics as they pass through the solar system (or impact Earth in the case of interstellar meteors), knowledge of the ISOs' present conditions can be tied to knowledge of their original birthplaces, provided they can be traced backwards in time to their home systems. 

Understanding the origins of interstellar asteroids and comets requires an examination of their dynamics within the context of our Milky Way galaxy. Our broader purposes here are 1) to bring the established principles of galactic stellar dynamics to bear on the question of the origin of these much smaller yet related bodies; and 2) tabulate a catalog of plausible galactic source regions with well-established positions and velocities, to serve as the basis for comparison for this and future studies of interstellar bodies. Here the Gaia DR 2 catalog \citep{gaiaDR2,Gaia16}, SIMBAD \citep{wenochegr00}, as well as the Catalog of Suspected Nearby Young Stars \citep{rieblulamcatalog18} and members of nearby moving groups compiled in \cite{gagmammal18} that contain full position and velocity information were used. 

This paper has three specific purposes. First, to quantitatively assess how far back in time we can trace the path of 'Oumuamua within the context of galactic dynamics. For this purpose we will model the Milky Way's gravitational potential, and the gravitational scattering processes which cause 'disk heating' ---the gradual increase in the velocity dispersion of disk stars over the course of time--- to assess their influence on the path of 'Oumuamua. The  phenomenon of disk heating is well documented and the age-velocity dispersion relation (AVR) has been well-established through the study of local stars. We determine that the path of our ISO can be traced back with some reliability perhaps as far back as 50 Myr or more. Secondly, the low velocity of 'Oumuamua with respect to the Local Standard of Rest (3-10 kms$^{-1}$, \cite{schbindeh10, cossakbil11}) hints that it might in fact be young. Adopting this as our working hypothesis, we model 'Oumuamua's path through the Milky Way to assess how close its origin point might be. If 'Oumuamua is indeed as young as its low peculiar velocity indicates, we find that its origin is likely to be currently within 1 kpc of the Earth, and thus available for detailed study if we can identify it. Thirdly, we attempt to identify the source region of 'Oumuamua (still under the assumption that it is young)  by modelling the path of 'Oumuamua and known astrophysically-likely regions in near-Sun space (YSOs, molecular clouds, etc.) to look for close approaches while accounting for the disk heating process. We do find particularly promising candidate source regions in the Carina and Columba moving groups, as well as the Lupus star-forming region (SFR). The discovery of comet 2I/Borisov\footnote{Minor Planet Electronic Circular 2019-S72 https://minorplanetcenter.net/mpec/K19/K19S72.html} during this work lead us to consider its motion within the Galaxy as well. However, its higher speed with respect to Local Standard of Rest indicates an older age, making tracing it back to its origin even more difficult. Nonetheless we performed a similar analysis on its past trajectory and report on some close encounters, though none are at low enough velocity to allow a compelling linkage.

\subsection{Possible sources of ISOs}

A number of possible sources of ISOs have been proposed in previous work, with recent developments inspired by the discovery of 'Oumuamua. Such investigations of ejection of material from planetary systems have taken place within a number of contexts. Planetesimals interacting with young planets \citep{ferip84}, the dynamics and formation of Oort Clouds and the comets and asteroids which constitute them \citep{dunquitre87,bramor13}, asteroidal and cometary ejecta from binary star systems \citep{jactamhamalirei18}, white dwarf tidal disruption events \citep{raf18}, young open clusters \citep{handehgra19}, and even ejection of smaller-scale dust from young main sequence stars, asymptotic giant branch (AGB) stars, and young stellar objects (YSOs) \citep{murweicap04} have been studied in some detail. 

Here we will not concern ourselves with the exact mechanism of ejection, except to note that, whatever the process, more energetic events are typically rarer than lesser ones. Thus ejection events just above the system's escape velocity are expected to be more common, and thus lower excess velocities are more likely. Thus a past encounter between 'Oumuamua and a particular astrophysical system at a low relative velocity will be considered more consistent with an origin there than an encounter at higher relative velocity, all else being equal. 

\subsection{Disk heating and 'Oumuamua's speed with respect to the Local Standard of Rest}

Stars form mainly in giant molecular clouds, which are confined more or less to the galactic plane with a scale height of about 80 pc and  a velocity dispersion $\approx 5$~kms$^{-1}$ \citep{jogost88,gamostjog91,fer01}. The distribution of stars in the Milky Way's disk is much broader (300~pc and 25-35 kms$^{-1}$ for the old thin disk near the Sun \citep{blager16}). This is attributed to 
'disk heating', the scattering of stars by molecular clouds or spiral arms as they travel together through the galactic potential. This process has been well studied both theoretically and observationally, though the relative contributions of giant molecular clouds (GMC) and spiral structure remain unclear (encounters with individual stars provide negligible scattering; \cite{bintre87}). Disk heating can be modelled rather well as a random walk with some initial velocity dispersion \citep{wie77}.
In particular, the phenomenon has a clear trend with age; young stars  have much lower deviations from the LSR than older populations (e.g. \cite{holnorand09}, Figure 7).

'Oumuamua would be subject to the same scattering processes as stars, and so the stellar age-velocity relation can be used to estimate its age. \cite{fenjon18} present an analysis of 'Oumuamua's velocity relative to the Local Standard of Rest (LSR). They note it deviates just 3 km~s$^{-1}$ from the LSR defined by \cite{schbindeh10}, and less than 10 km~s$^{-1}$ from that defined by \cite{cossakbil11}.  These values are lower than are usually covered by galactic age-velocity relations, but imply an age less than 1 Gyr; for example, an extrapolation of \citep{holnorand09} Figure 7 yields an age $\sim 100$~Myr, while \citep{robbiefer17} reports an age less than 150~Myr for the sub-component corresponding to 'Oumuamua's velocity. Other authors have also argued 'Oumuamua's small deviation from the LSR indicates it has undergone only a small amount of dynamical perturbation throughout its journey \citep{gaiwilkra17}. 

'Oumuamua's low velocity with respect to the LSR is a clue that it may be young, though not proof; it could indeed be as old as the Milky Way itself, but the probability of drawing a value less than or equal to 10 kms$^{-1}$ from a Maxwell-Boltzmann distribution with a most probable value of 30 kms$^{-1}$ is only 0.026. On the other hand, \cite{gai18} note that given the average age of disk stars is a few Gyr, observing an exceptionally young object presents a puzzle. Whether ejected during the planet formation stage or by early interactions between cluster stars \citep{handehgra19}, most ISOs should be a few Gyr in age. \cite{gai18} mention that strong selection effects against an older population would have to account for our observation of such a young ISO as 'Oumuamua. However, the discovery of 2I/Borisov, if it is indeed older as is suggested by its higher speed with respect to the LSR, casts doubt on the existence of such selection effects. Though 'Oumuamua could share the motion of a young cluster by coincidence, \cite{gai18} report that the probability of 'Oumuamua sharing the same kinematics as a moving group $\lesssim$ 100 Myr old by coincidence is $<$ 1 percent. This is in line with \cite{rieblulam18}'s determination that the ratio of field stars (young and old) to young stars belonging to moving groups within 25 pc is 50:1. These simple arguments imply that, though observing a young ISO is a priori unlikely, this does not mean 'Oumuamua is unlikely to be young, given the additional information provided by its kinematics.

In the absence of any other information we will adopt and examine here the hypothesis that 'Oumuamua is young (less than 100~Myr old).

We note the useful fact that 1~km~s$^{-1}$ $\approx$ 1~pc~Myr$^{-1}$. This allows us to quickly estimate that if 'Oumuamua was ejected at low speed (say 1 kms$^{-1}$) from its birth system 100 Myr ago, 'Oumuamua has travelled only about 100 pc since it was ejected, or in other words, the source of 'Oumuamua is 100 pc away from the Earth currently (since 'Oumuamua is nearby). Of course, during that 100 Myr both 'Oumuamua and its home system will have completed 40\% of an orbit around the Galaxy, travelling about 20,000~pc as they do so, but their displacement relative to each other is much less.  Though this simple analysis has not yet accounted for the random walk of disk heating, the two reasonable assumptions that 1) 'Oumuamua's low velocity with respect to the LSR implies it may be young and 2) that low ($\las 1$~kms$^{-1}$) ejection speed processes are more efficient than high-speed ones, lead us to examine a possible origin for 'Oumuamua in the local galactic neighbourhood.

\section{Model}
 
\subsection{Our model of the Milky Way Galaxy}

We treat the gravitational field of the galaxy using a time-independent, three-component function as developed by \cite{miynag75}. Accounting for contributions from the galactic disk, bulge, and halo produces, in galactocentric cylindrical coordinates,

\begin{eqnarray}
&\Phi_{gal} = \Phi_{d} + \Phi_{b} + \Phi_{h},\\
&\Phi_{gal} = -\sum\limits_{i=d,b,h}\frac{GM_{i}}{\sqrt{R^{2} + \big(a_{i} + \sqrt{z^{2} + b_{i}^{2}}\big)^{2}}},
\end{eqnarray}

as widely used in previous studies (for example \cite{baifarmee18} and \cite{zulsansuc18}). The length scales $a$ and $b$ are tuned to reflect the geometries of the galactic disk, bulge, and halo components. Setting $b\ll a$ we recover a Kuzmin potential which we use to model the flat geometry of the disk, while setting $a\ll b$ produces a Plummer sphere which models the spherical bulge and halo components. We take the values for $a$, $b$ and the mass components for each geometry from \cite{daucol95}. Although newer mass estimates have been made (\cite{porwegger15} estimate a portion of the bulge mass, and \cite{blager16} record estimates for the dark matter and disk masses, as well as characteristic lengths for the various geometries), we estimate that those of \cite{daucol95} are generally within a factor of two of these newer values. Changing them does not significantly affect the results of integrations through this potential. The constants used are summarized in Table \ref{tab:galactic_parameters}. 

All coordinate transformations are done using {\tt\string Astropy}, version 3.1.2. Astrometry for all stars and 'Oumuamua is instantiated in the ICRS frame, and transformed to a galactocentric cylindrical frame via the {\tt\string transform$\_$to} and {\tt\string represent$\_$as} methods before numerical integration. In defining its Galactocentric frame, {\tt\string Astropy} derives the galactic center position from \cite{reibru04}, the solar distance from galactic center from \cite{gileistri09}, the sun's height above the galactic plane from \cite{chestosmi01}, the solar peculiar velocity from \cite{schbindeh10}, and the circular velocity at the solar radius from \cite{bov15}. These constants are also included in Table \ref{tab:galactic_parameters}. 

\begin{deluxetable*}{cCCCC}[h]
\tablecaption{Parameters Defining Galactic Geometry and {\tt\string Astropy}'s Galactocentric Frame. \label{tab:galactic_parameters}}
\tablecolumns{4}
\tablenum{1}
\tablewidth{0pt}
\tablehead{
\colhead{Parameter} &
\colhead{Units} &
\colhead{Value} & 
\colhead{Reference}}
\startdata
a$_{i}$, i=d, b, h & [pc] & 3500, 0, 0 & (1) \\
b$_{i}$, i=d, b, h & [pc] & 250, 350, 24 000 & (1) \\
M$_{i}$, i=d, b, h & [10$^{10}$M$_{\odot}$] & 7.91, 1.4, 69.8 & (1) \\
Galactic center position (RA, Dec) &  [hr/min/s, deg/'/''] & 17:45:37.224, -28:56:10.23 & (2) \\
Solar distance from center & [kpc] & 8.33 $\pm$ 0.35 & (3) \\
Solar height above galactic plane & [pc] & 27 $\pm$ 4 & (4) \\
Solar peculiar velocity (U,V,W) & [kms$^{-1}$] & 11.1$^{+0.69}_{-0.75}$,12.24$^{+0.47}_{-0.47}$,7.25$^{+0.37}_{-0.36}$ & (5) \\
Circular velocity at solar radius & [kms$^{-1}$] & 218 $\pm$ 6 & (6) \\
\enddata
\textbf{References}.(1) \cite{daucol95}, (2) \cite{reibru04}, (3) \cite{gileistri09}, (4) \cite{chestosmi01}, (5) \cite{schbindeh10}, (6) \cite{bov15}.
\end{deluxetable*}

\subsubsection{Initial Conditions of 'Oumuamua} \label{initialconditions}

We use two separate solutions for the orbit of 'Oumuamua. The asymptotic initial conditions (at time -infinity) of 'Oumuamua are provided by \cite{baifarmee18}, in which they numerically integrated its orbit back to -3000 BCE, then extrapolated a Keplerian orbit back to time -infinity. Throughout our determination of possible source regions of 'Oumuamua in section 3.3, we characterize the uncertainty in their 2k=2 solution for 'Oumuamua's orbit by generating 100 clones of 'Oumuamua, each having an initial velocity randomly sampled from a distribution whose covariance matrix  is recorded in \cite{baifarmee18}. We determine our own solution, by creating clones derived from the most recent JPL orbit and covariance matrix\footnote{https://ssd.jpl.nasa.gov/sbdb.cgi, retrieved 18 June 2019} and back-integrated with the RADAU \citep{eve85} algorithm with an error tolerance of $10^{-12}$, under the influence of the Sun and the planets. Our own solution is recorded in Table \ref{tab:initial_conditions}, while the 2k=2 solution along with its covariance matrix are recorded in \cite{baifarmee18}. 

\begin{deluxetable*}{cCCCCCCCCCCCC}
\tablecaption{Our solution for the initial conditions of 'Oumuamua in the ICRS frame, at -10 000 years. The asymptotic 2k=2 solution provided by \cite{baifarmee18} was also used in this study; only our own determination is supplied here. Coordinates are recorded in Right Ascension ($\alpha$), Declination ($\delta$), distance ($d$), radial velocity ($v_{r}$), and the proper motions in Right Ascension and
Declination are $\mu_{\alpha}$ (corrected for $\cos\delta$), and $\mu_{\delta}$, respectively. Each coordinate's 1-$\sigma$ uncertainties are provided. \cite{baifarmee18} provide the 2k=2 solution complete with its covariance matrix. \label{tab:initial_conditions}}
\tablenum{2}
\tablewidth{0pt}
\tablehead{
\colhead{$\alpha$} & \colhead{$\sigma_{\alpha}$} &
\colhead{$\delta$} & \colhead{$\sigma_{\delta}$} & 
\colhead{$d$} & \colhead{$\sigma_{d}$} & 
\colhead{$v_r$} & \colhead{$\sigma_{v_r}$} & 
\colhead{$\mu_{\alpha}$} & \colhead{$\sigma_{\mu_{\alpha}}$} & 
\colhead{$\mu_{\delta}$} & \colhead{$\sigma_{\mu_{\delta}}$} \\
\colhead{[deg]} & \colhead{[deg]} & 
\colhead{[deg]} & \colhead{[deg]} &
\colhead{[AU]} & \colhead{[AU]} &
\colhead{[kms$^{-1}$]} & \colhead{[kms$^{-1}$]} &
\colhead{[masyr$^{-1}$]} & \colhead{[masyr$^{-1}$]} &
\colhead{[masyr$^{-1}$]} & \colhead{[masyr$^{-1}$]}
}
\startdata
279.5588 & 0.002746 & 33.8795 & 0.001484 & 55674.1550 & 2.0930 & -26.4052 & 0.003111 & -0.3249 & 1.4432 & 0.02967 & 1.8027 \\
\enddata
\end{deluxetable*}

\subsection{Integrators}

We will report on two types of simulations here. The first involves the back-integration of 'Oumuamua alone within the potential of our Galaxy under the effect of disk heating. This is to examine the effects of gravitational scattering on the asteroid's past trajectory and assess how far back we can expect to go reliably. These simulations are performed with the RADAU \citep{eve85} integrator in Cartesian coordinates with a output time step of 1000 years and an error tolerance of $10^{-12}$. 

The second set of simulations computes the back-trajectory of 'Oumuamua and many stars and other systems, and includes the galactic potential but no random impulses. These simulations are designed to examine close encounters between the nominal paths of known galactic objects and 'Oumuamua to assess possible origin points for this asteroid. For these simulations, {\tt\string SciPy}'s {\tt\string solve$\_$ivp} {\tt\string RK45} implementation of a mixed fourth/fifth order Runge-Kutta integrator method is used in numerically integrating the motion of the objects. This method is called to repeatedly integrate the positions and speeds of objects every 10 000 years. In each 10 000 year interval the distance to 'Oumuamua from catalog objects is checked before integrating backward another 10 000 years. The time step it takes over each 10 000 year interval is set automatically by the solver. The error tolerances are the default values, 10$^{-6}$ in absolute tolerance and 10$^{-3}$ in relative tolerance. The integrations are carried out in cylindrical coordinates according to the equations of motion in a cylindrical system under a potential,

\begin{eqnarray}
&\ddot{R} = -\frac{\partial{\Phi}}{\partial R} + R\dot{\theta}^{2},\\
&\ddot{\theta} = -\frac{1}{R^{2}}\frac{\partial{\Phi}}{\partial \theta} - 2\frac{\dot{R}\dot{\theta}}{R},\\
&\ddot{z} = -\frac{\partial{\Phi}}{\partial{z}}.
\end{eqnarray}

\subsection{Disk heating models}

For simulations that include the effects of disk heating, that effect will be modelled as a series of random kicks to the velocity of each of an ensemble of particles ('clones') with similar initial conditions as we integrate them backwards within the Galactic potential. The increasing dispersion in the positions and velocities of this ensemble as we go further into the past will provide a measure of how well we can hope to locate the true position of 'Oumuamua at that time. 

Velocity kicks of 10 kms$^{-1}$ are applied independently to each clone at random times according to a Poisson process with a characteristic time scale of 200~Myr; these have been found to provide the best fit to the local galactic age-velocity relation \citep{wie77, mihbin81}. Because the precise nature of the deflection depends on the details of the encounters, which are not known, we apply the kicks in three ways to examine the envelope of possible outcomes. 
\begin{enumerate} 
\item A velocity kick of 10 kms$^{-1}$ applied in a random direction on the sphere ('fixed kick')
\item A velocity kick drawn from a three dimensional Maxwell-Boltzmann distribution with a most probable speed of 10 kms$^{-1}$ ('Maxwellian')
\item A deflection of the velocity vector without changing its magnitude ('Maxwellian deflection'). In this case,  kicks drawn from a one-dimensional Maxwell-Boltzmann distribution with most probable speed of 10 kms$^{-1}$ are applied along two orthogonal directions perpendicular to the velocity, which is then renormalized to its original length. This corresponds to the rotation of the velocity vector expected from long-range 'dispersion-dominated' encounters \citep{bintre87}.
\end{enumerate}

\section{Results and discussion}

\subsection{Integrations with disk heating: past trajectory} \label{diskheating}

\begin{figure}[ht!]
\plottwo{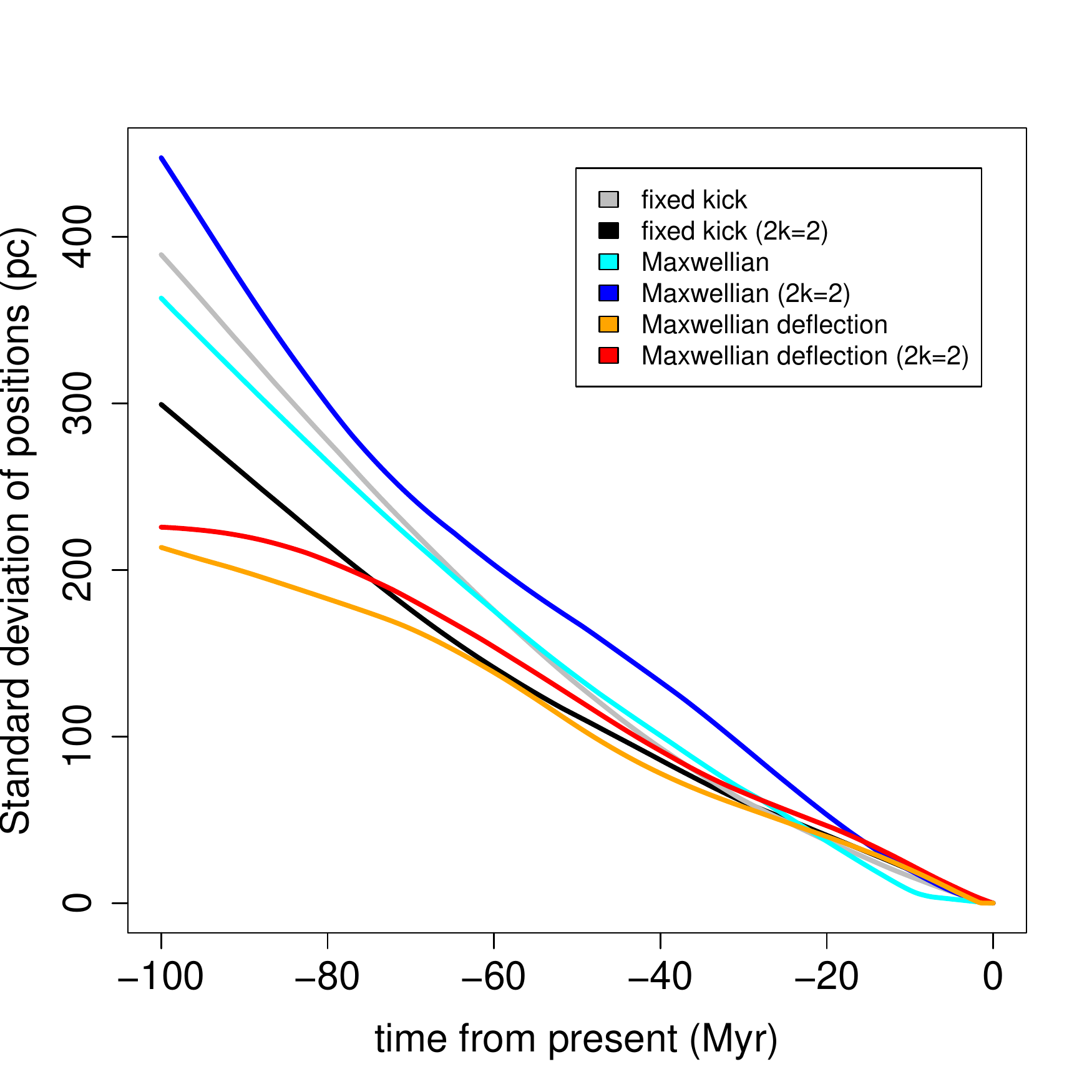}{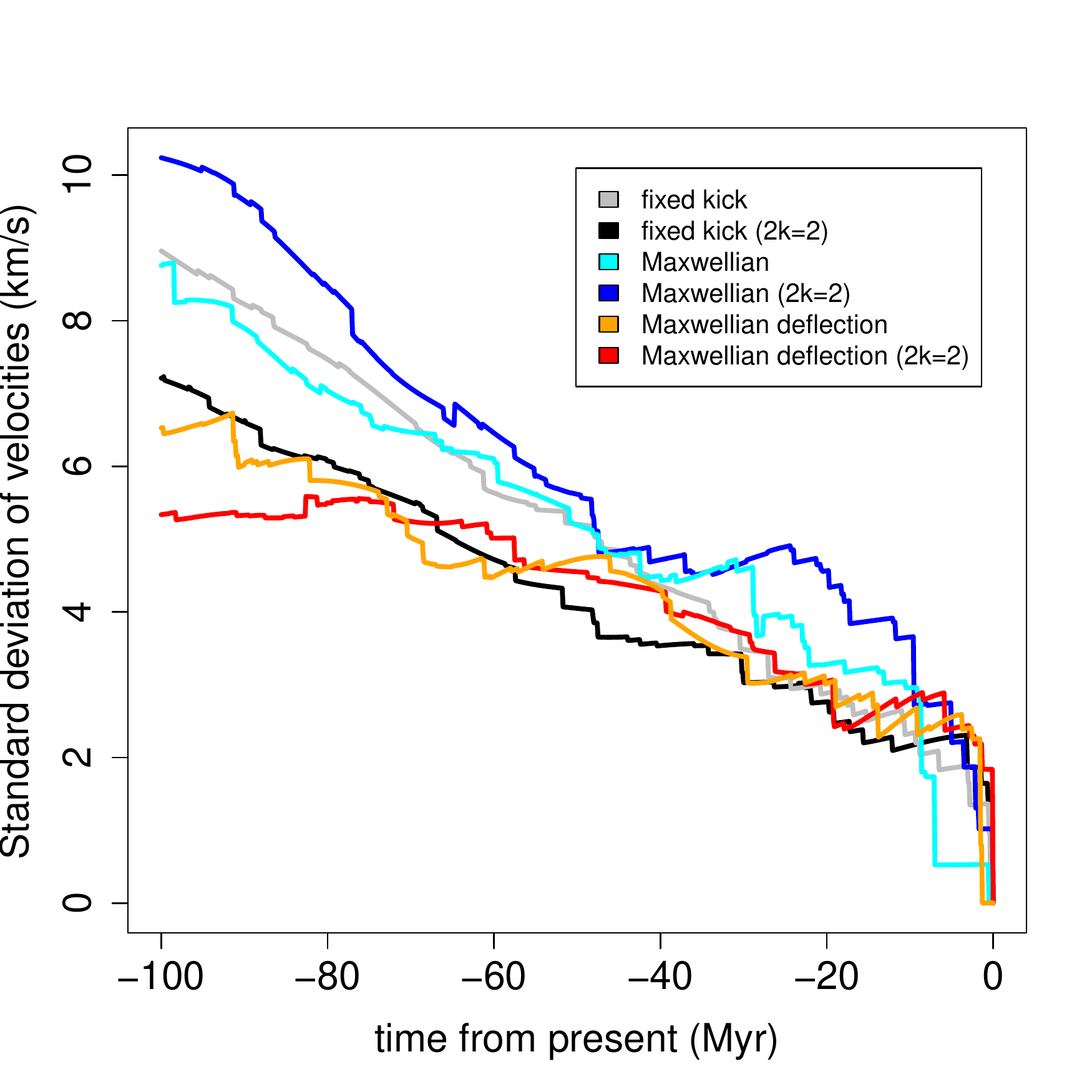}
\caption{The standard deviation of the position and velocity of clones of 'Oumuamua integrated backwards within the Galaxy. \label{fi:dispersion}}
\end{figure}

The increase in the dispersion, as measured by the standard deviation, of the positions and velocities of our clones of 'Oumuamua are presented in Figure~\ref{fi:dispersion}. Two sets of clones are examined, based on our own determination of the pre-arrival velocity of 'Oumuamua and the 2k=2 solution from \cite{baifarmee18}, which is based on the \cite{micfarmee18} solution with pure radial acceleration proportional to heliocentric distance $r^{-2}$.
As we go back in time, the dispersion of the clones grows, reaching up to roughly 15 pc and 2 kms$^{-1}$ at 10 Myr, 100 pc and 5 kms$^{-1}$ at -50 Myr, and 400 pc and 10 kms$^{-1}$ at -100 Myr, depending on the model chosen.   Recall that at this point we are not integrating any stars or other objects within our Galaxy; we are purely assessing the uncertainty in the position and velocity of 'Oumuamua.

As we investigate possible origin systems for 'Oumuamua, Figure \ref{fi:dispersion} will serve as a guide as to how close an encounter and how low a relative speed constitute an interesting encounter in terms of a possible origin. Though we cannot know what impulses 'Oumuamua has undergone during its past, we can state that the true position and velocity of 'Oumuamua at some specific time in the past is most likely within the dispersion given by our models from its nominal trajectory. As a result, the source system of 'Oumuamua is as well, and we will use this to try to constrain its location of origin. Thus, Figure \ref{fi:dispersion} provides us with a 'sphere of interest' around 'Oumuamua at any point in time.

It is worth noting that the growing 'sphere of interest' around 'Oumuamua as we move into the past creates a clear danger that mere coincidence might provide an erroneous good match with a candidate system of origin. A close encounter between 'Oumuamua and a potential source region within our Galaxy 50 Myr ago at a relative speed of 5 kms$^{-1}$ and a distance of 100~pc would provide roughly as good a match as one 10~Myr ago at 15~pc and 2~kms$^{-1}$.  We will try to avoid this trap by carefully discussing the context of any interesting origin candidates found, and in any case at least, Figure \ref{fi:dispersion} provides us with a quantitative measure of the degree of danger at any point.

\subsection{Integrations with impulses: distance to 'Oumuamua's origin point}

The end states of our ensemble of the previous section provide a probability distribution outlining our uncertainty in the location of 'Oumuamua up to 100 Myr ago. If that young age is correct, and 'Oumuamua was ejected at a low relative velocity, then the origin system of 'Oumuamua was, from Figure \ref{fi:dispersion}, within about 400 pc of our mean end state 100 Myr ago. In order to assess where that origin point might be currently, we will use those afore-mentioned end states as proxies for the origin point's location 100 Myr ago, and integrate them forward to the present day. Such a use of 'Oumuamua's end states to model the rough dynamics of its origin system will be correct as long as 1) 'Oumuamua and its point of origin were co-located at 'Oumuamua's time of origin (satisfied by definition) and 2) they were travelling at low relative velocity at the time of ejection (which is the most likely case for most ejection mechanisms). 

\begin{figure}[ht!]
\plotone{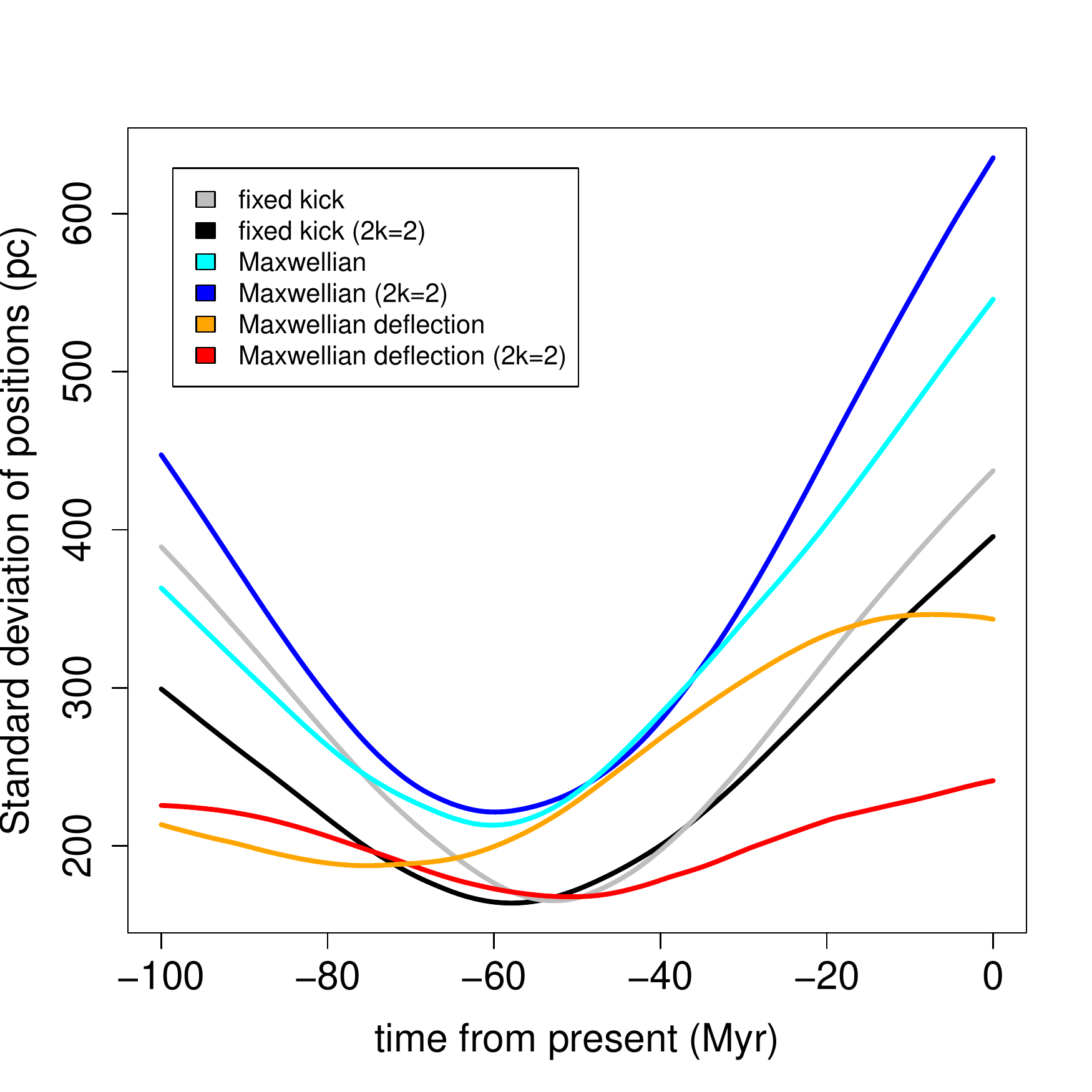}
\caption{The dispersion of the modelled origin system positions (relative to the mean at a given time) integrated forwards to the current time within the Galaxy.  At -100 Myr, there is an initial spread in position of the origin systems, reflecting the spread in where 'Oumuamua may have been 100 Myr ago (as calculated in section 3.1 Fig~\ref{fi:dispersion}). We assume that the origin system has the same position and velocity as 'Oumuamua at -100~Myr, and then integrate these systems forward under disk heating. The dispersion at the right hand side of the plot ($\sim 500$~pc) represents the envelope of how far the origin system could be from 'Oumuamua (and for all practical purposes, Earth) at the present day. \label{fi:dispersion-forw}}
\end{figure}

Figure~\ref{fi:dispersion-forw} shows the results of these forward simulations. The position and velocity dispersions start with decreasing trends as one moves left to right as the previously expanding cone of 'Oumuamua clones initially contracts. Eventually, the impulses due to disk heating cause enough scatter for them to begin diverging again, which takes about 40~Myr. The standard deviation (1-$\sigma$) in position and velocity at the right-hand side of both panels of Figure ~\ref{fi:dispersion-forw} imply that the origin system of 'Oumuamua is at this moment within 300-600 pc. At the 2-$\sigma$ or 95\% confidence level, 'Oumuamua's origin system should be within 1.2 kpc of Earth, within the local Orion Arm, and thus relatively easily accessible to Earth-based study. There are a number of nearby star-forming regions and moving groups, plausible sources of asteroids, such as Perseus, Orion, Taurus and Ophiuchus which we will assess as possible origin points in the next sections. 

\subsection{Integrations without impulses: close encounters}

\subsubsection{Extending previous results}

A number of efforts have been made to understand 'Oumuamua in the context of the close galactic neighbourhood, including but not limited to the work of \cite{mam17}, \cite{gaiwilkra17}. \cite{baifarmee18}, \cite{zulsansuc18}, \cite{fenjon18}, \cite{portorpel18}, \cite{dybkro18}, \cite{gai18}, \cite{zha18} and \cite{eub19}.  Here we extend these by providing a uniform treatment of all stars in the Gaia DR2 catalog, the Catalog of Suspected Nearby Young Stars \citep{rieblulamcatalog18}, the compilation of bonafide members of kinematic groups in \cite{gagmammal18}, and SIMBAD, under an understanding of the limitations imposed by disk heating, as derived in earlier sections.

As other authors have pointed out, given that 'Oumuamua is likely young, it may have come from one of the many sources of young stars nearby in the galaxy. Young stars are most frequently observed as members of star-forming regions (SFRs), open  clusters, and moving groups. Some of the closest SFRs include Scorpius-Centaurus (Sco-Cen) at $\sim$ 118-145 pc \citep{premam08}, Taurus-Auriga at $\sim$ 140 pc \citep{galloiort18}, Chamaeleon I, II, and III at $\sim$ 180-200 pc \citep{voimanpru18}, and the Lupus clouds at $\sim$ 110-190 \citep{galbertei13}. 

Many young stars within 100 pc of the Sun are categorized into members of open clusters and nearby young moving groups (NYMGs) \citep{rieblulam18}. Clusters and moving groups are thought to be born in molecular clouds, and slowly disperse over millions of years due to feedback between the stars and material left over from the star formation process \citep{wrimam18}. Before dispersing into the galaxy, the young stars form an \textit{association}, a group of stars which share similar kinematics. A number of kinematically-related young stars have been identified in the past few decades, and the associations and moving groups they constitute have been continuously redefined with the advent of better astrometric surveys. The most up-to-date classifications define three open clusters within 100 pc of the sun (Coma Ber, Hyades, and $\eta$ Cha), and ten NYMGs \citep{rieblulam18}.

\cite{rieblulam18} provide a thorough overview of the current understanding of the open clusters and moving groups within 100 pc of the sun. Moving groups are distinguished from open clusters by the fact that they are loose and gravitationally unbound. Moving groups are older than SFRs, but much closer. All stars contained within a moving group or open cluster are thought to originate from the same star forming event, and so share the same position in space at their time of formation, as well as the same age, composition, and because they are young, the same galactic space motion as their natal gas clouds \citep{rieblulam18}.

Of the previous studies which mention moving groups or kinematic associations in the context of 'Oumuamua (such as \cite{fenjon18}, \cite{gai18}, \cite{gaiwilkra17}, or \cite{eub19}), \cite{fenjon18} include integrations of the motion of 'Oumuamua and stars in the Pleiades association, \cite{gai18} present integrations of 'Oumuamua along with the mean positions and velocities of many of the nearest moving groups (including the Carina and Columba Associations), while \cite{gaiwilkra17} and \cite{eub19} only consider the UVW kinematics of the moving groups with respect to 'Oumuamua without integrations. \cite{gaiwilkra17} in particular point out the close kinematic match between 'Oumuamua and the Carina and Columba associations, finding speed deviations $\lesssim$ 2 kms$^{-1}$. They suggest that 'Oumuamua was ejected from a protoplanetary disk in these groups $\approx$ 40 Myr ago. We extend these results by explicitly integrating the known and suspected members of the NYMGs and open clusters within 150 pc, along with other stars such as YSOs which belong to the nearest SFRs. \cite{gai18} report that integrating the mean positions and velocities of the moving groups was found not to reveal any close encounters between them and 'Oumuamua in the past. However, we find many individual slow encounters with members of these groups which are close enough to plausibly interact with 'Oumuamua if dynamical heating is accounted for. The NYMGs which have the most such encounters are Carina and Columba. The stars in Carina and Columba minimize their distance to 'Oumuamua roughly during the moving groups' formation epoch, $\sim$30-45 Myr in the past (\cite{rieblulam18} and references therein). Since we expect bodies like 'Oumuamua to be ejected in large numbers and at low velocities during the stages of planet formation, as well as during the early, tightly-bound stages of moving group evolution due to intra-cluster interactions \citep{handehgra19}, the low encounter distances and velocities observed with respect to the Carina/Columba association just as it is forming make it a particularly attractive candidate source region.

We draw stars which are bonafide or suspected members of each moving group from two sources. The principle source we use is the Catalog of Suspected Nearby Young Stars as supplied in \cite{rieblulamcatalog18}. This catalog is intended as 'a single resource for studying the individual and ensemble properties of young stars' \citep{rieblulam18}. It has been compiled from careful scrutiny of numerous publications of moving group candidates, and contains basic information for any star system within 100 pc that has ever been reported as young. Of the many properties of the young stars compiled, their astrometry and moving group membership are recorded. We use each star's \textit{GROUP1} entry in the catalog to categorize it into each NYMG, and regard the stars with positive \textit{Bonafide} entries as high-confidence members. The catalog supplies astrometry for 4 727 stars with \textit{GROUP1} classifications.  For additional data, we supplement with bonafide members compiled from the literature in \cite{gagmammal18} (Table 5). The populations of suspected and bonafide moving group members recorded in the Catalog of Suspected Nearby Young Stars is provided in Table \ref{tab:riedel_groups}, while the number of bonafide stars in each group recorded in \cite{gagmammal18} are given in Table \ref{tab:gagne_groups}.

We further supplement these data with all stars from SIMBAD with full astrometric solutions recorded, classified by object type (Young Stellar Objects, white and brown dwarfs, low-mass stars, etc.). We also use all Gaia Data Release 2 stars with full astrometric solutions recorded. Although many of the Gaia and SIMBAD stars have previously been considered (\cite{baifarmee18}, \cite{dybkro18}, \cite{zulsansuc18}, \cite{fenjon18}, \cite{portorpel18}, etc.), we check our results against these previous studies' and extend them by looking specifically for YSOs in SFRs, as well as stars in nebulae and other extended objects.

In addition to integrating 100 clones of 'Oumuamua from the 2k=2 solution, we sample the astrometric coordinates of the stars in our data set from Gaussian distributions with means and standard deviations given by the nominal values and their corresponding uncertainties. Altogether 100 clones of 'Oumuamua interact with 100 clones of each star in each of our integrations to provide distributions in the relative speeds and distances at the time of closest approach. We use the 2k=2 solution for consistency with previous studies. However, we find that in general the uncertainty in 'Oumuamua's initial conditions is dwarfed by the uncertainties in the astrometric data used in our integrations; our results are not strongly sensitive to whether the 2k=2 or our own clones of 'Oumuamua are used. The results differ at the $\sim$ 100 ms$^{-1}$ level in speed, and at the $\sim$ 0.1 pc level in median encounter distance.

Although the Catalog of Suspected Nearby Young Stars may be considered complete up to January 2015 \citep{rieblulam18}, newer astrometry exists for many of the stars due to Gaia. However, no database has been compiled that contains the same membership data as the Catalog of Young Stars but with the newest astrometry. We checked our results by re-integrating each star in Table \ref{tab:candidate_table} for which we used the Catalog of Young Stars with all available Gaia measurements (excluding Gaia radial velocities with higher uncertainties than those recorded in the Catalog). We find significantly different results for a minority of stars; two of the 16 Carina stars were found not to pass close enough to 'Oumuamua, and one bonafide member, HD 83096, had an encounter time of 16.53 Myr as opposed to the 31.02 Myr recorded in Table \ref{tab:candidate_table}. Nevertheless, the average time of encounter between the 14 Carina stars was -31.82 Myr, and between the four Bonafide stars it was -28.62 Myr. The average time of closest approach between all 21 Columba stars in Table \ref{tab:candidate_table} was found to be -26.00 Myr in the past, while the average encounter time between the 9 bonafide members was -24.73 Myr. We also find two additional Carina stars (HD 44345 and TYC 8927-2869-1) from the Catalog not included in Table \ref{tab:candidate_table}, for which the Gaia data produces consistent encounters, while the Catalog's data does not (median distances and speeds of 22.95 and 53.19 pc, 5.69 and 5.06 kms$^{-1}$, at -29.69 and -35.20 Myr, respectively). The results are therefore only weakly sensitive to whether the Gaia astrometry or that contained in the Catalog of Young Stars is used. An illustration of the approach of 'Oumuamua to Earth is shown in Animated Figure~\ref{Fig:flythrough_borisov}.

\begin{deluxetable*}{cCCC}[h]
\tablecaption{Suspected and bonafide members of moving groups from \cite{rieblulam18} considered in this work. Each total is the number of stars in the group with radial velocity data recorded. Group classification is based on each star's \textit{GROUP1} entry in the catalog. \label{tab:riedel_groups}}
\tablecolumns{4}
\tablenum{3}
\tablewidth{0pt}
\tablehead{
\colhead{Group} &
\colhead{No. total} &
\colhead{No. Bonafide}
}
\startdata
Columba & 103 & 15 \\
Carina & 38 & 3 \\
TW Hydra & 77 & 17 \\
AB Dor & 338 & 34 \\
Pleiades/Local Association & 78 & 16 \\
Coma Ber & 54 & 33 \\
Tucana-Horologium & 311 & 32 \\
Ursa Major & 271 & 0 \\
$\eta$ Cha & 15 & 0 \\
Hercules-Lyra & 53 & 2 \\
Beta Pic & 216 & 28 \\
IC 2391 & 37 & 0 \\
Hyades & 12 & 0 \\
\enddata
\end{deluxetable*}

\begin{deluxetable*}{cCCC}[h]
\tablecaption{Bonafide Members of Moving Groups From \cite{gagmammal18} Considered in This Work \label{tab:gagne_groups}}
\tablecolumns{4}
\tablenum{4}
\tablewidth{0pt}
\tablehead{
\colhead{Group} &
\colhead{No. Total (=Bonafide)}
}
\startdata
Columba & 19 \\
Carina & 5 \\
TW Hydra & 22 \\
AB Dor & 47 \\
Pleiades/Local Association & 187 \\
Coma Ber & 41 \\
Tucana-Horologium & 31 \\
$\eta$ Cha & 16 \\
Beta Pic & 47 \\
IC 2391 & 10 \\
Hyades & 165 \\
Carina-Near & 18 \\
Corona Australis & 13 \\
$\epsilon$ Cha & 23 \\
IC 2602 & 10 \\
Lower Centaurus Crux & 81 \\
Octans & 10 \\
Platais 8 & 11 \\
$\rho$ Ophiuchi & 180 \\
Taurus & 121 \\
32 Orionis & 31 \\
Upper Centaurus Lupus & 102 \\
Upper CrA & 23 \\
Ursa Major cluster & 7 \\
$\chi^{1}$ For & 11 \\
\enddata
\end{deluxetable*}

\subsubsection{Candidate Regions}

The systems which move at appropriately low speed (we take our speed bound to be $\lessapprox$ 8 kms$^{-1}$) and whose distances from 'Oumuamua are within the bounds imposed by disk heating at the time of encounter are summarized in Table \ref{tab:candidate_table}, along with each system's object type, time of encounter, and membership in any SFR or moving group. Bonafide members of moving groups are labelled with an additional '$^{*}$' beside their identifier. The table does not follow any inherent ranking of the candidates.

The most interesting candidate source regions are those with the most encounters consistent with a low relative speed ($\lessapprox$ 8 kms$^{-1}$), encounter distances within the bounds imposed by disk heating, and those for which the closest encounters occur at times that are allowed considering the age of the group, SFR, or individual star. The most interesting source regions we find are the Carina and Columba moving groups. 'Oumuamua has its closest encounter to multiple members of these groups at $\lesssim$ 5 kms$^{-1}$, at median times of -32.89 Myr and -25.63 Myr, respectively, close to the 30-45 Myr age range of these systems \citep{rieblulam18}, when they can be expected to be ejecting the most material. Another strong case can be made for the Lupus SFR. We find four encounters with YSOs at $5-7$ kms$^{-1}$. Two are members constituting the Lupus core moving group \citep{galbertei13}, and two have been identified as members of the Lupus population of weak line T-Tau stars \citep{galbertei13}, the ages of which have been found to be considerably older than the Lupus classical T-Tau population \citep{mak07}. To explain this spread in ages it has been suggested that there may have been multiple bursts of star formation events in Lupus \citep{mak07} and this would roughly align with the epoch of 'Oumuamua's closest approach to these two stars. We find a promising encounter with the T-Tau star V391 Ori, possibly associated with the 32 Orionis moving group, moving at just $3.84$ kms$^{-1}$ relative to 'Oumuamua. A Carina star, GJ 1167 A is a young M dwarf passing within a distance close to the disk heating bounds, 2 Myr ago, at just 4.9 kms$^{-1}$. Other YSOs with plausible but less compelling encounters have been identified belonging to the Taurus-Auriga SFR. Two of these encounters are at times when the SFR may have been active. A detailed analysis of our results is provided below.

\begin{itemize}

\item{\textbf{Carina and Columba moving groups}:} A number of bonafide or suspected members of the Carina and Columba moving groups are recorded in Table \ref{tab:candidate_table}. Of the 16 stars associated with Carina in Table \ref{tab:candidate_table}, four are bonafide members according to \cite{rieblulam18} and \cite{gagmammal18} (HD 49855, HD 55279, HD 83096 a and b, and V479 Car). HD 83906 a and b make up one multiple star system. Their encounter conditions are similar, so we take the system's encounter with 'Oumuamua to be that of HD 83096 a. Depending on the exact model used for disk heating, the spread in distance between clones of 'Oumuamua at -30 Myr is $\sim$ 58-68 pc, and at -35 Myr is $\sim$ 68-80 pc (see Figure \ref{fi:dispersion}, where further in the past the 'Maxwellian' model gives larger dispersions than the 'Maxwellian deflection'); most candidates listed in Table \ref{tab:candidate_table} fall well within these bounds. There are in total 6 unique bonafide members of Carina in our initial data set. The age of Carina has been estimated between 30 and 45 Myr (see \cite{rieblulam18} and references therein). The median encounter time between the bonafide Carina stars is -31.46 Myr, and for all Carina stars in Table \ref{tab:candidate_table} the median encounter time is -32.89 Myr. 'Oumuamua minimizes its distance from a number ($\sim 40\%$) of Carina stars, (15 out of 41 total), four of which are bonafide members out of six total, at the minimum age of the group. This suggests 'Oumuamua may have had an encounter with the group at a very early stage in its history, either during the initial star-forming epoch or after, as the young stars gradually dispersed.

One star in particular in Carina, GJ 1167 A, passes within 3.94 pc (2.41-11.64 pc in the 5th and 95th percentiles), at 4.90 kms$^{-1}$ (2.0-8.06 kms$^{-1}$ in the 5th and 95th percentiles), 2.37 Myr ago. The dispersion due to disk heating at 2.37 Myr is $\sim$ 3.6 pc. GJ 1167 A and B is a binary system (though is not likely a physical binary, see for example \cite{statanbry10}, or \cite{bowliushk15}). Currently $\sim$ 12 pc away, it is composed of two young M dwarfs, and therefore likely less than 1 M$_{\odot}$ combined. Our own Oort cloud extends only 1 pc from our star \citep{donweilev04} so the tidal radius of GJ 1167 is likely much smaller than this. Its low speed and distance close to the disk heating bounds do make this encounter more compelling than many of the encounters with smaller distances but much higher speeds reported in previous works, though its small tidal radius and relatively large spread in encounter distance mean it is not likely the home of 'Oumuamua.

Of the 21 stars associated with Columba in Table \ref{tab:candidate_table}, nine are bonafide members; HD 40216, HD 38206, HD 30447, HR 8799, HIP 1134, HD 32372, HD 32309, HD 37484, and 2MASS J05184616-2756457. There are in total 29 unique bonafide Columba stars in our initial data set. The median time of encounter between the 9 bonafide stars is 23.85 Myr ago, while the median encounter time between all Columba stars in Table \ref{tab:candidate_table} is 25.63 Myr. Disk heating gives a spread of $\sim$ 46-54 pc at this time. The estimated age of the Columba group is between 30-42 Myr in the past \citep{rieblulam18}. 'Oumuamua therefore minimizes its distance to a lesser number ($\sim$ 18$\%$) of Columba stars, slightly before the group's minimum age.

Given that there are multiple groups with more members than Carina and/or Columba (Pleiades, Tucana-Horologium, Ursa Major, and Beta Pictoris, some of which will be detailed in the following sections), these two groups nevertheless produce the largest number of reasonable encounters (ie. with low-speed and distances within the disk heating bounds), indirectly supporting an origin in this group.

\item{\textbf{Lupus SFR}:}
The age of the Lupus SFR has been estimated to be ~3 Myr \citep{hugharkra94}, but newer analyses have found a considerable spread in ages of the stars in Lupus (eg. \cite{mak07}). The population of weak-line T-Tau stars were found to be older, several of them being $\sim$25-30 Myr old, while half the classical T-Tau stars were found to be less than 1 Myr old (see their Figure 4). At least two of our Lupus candidates are indeed weak-line T-Tau stars. \cite{mak07} note that a possible explanation for the large spread in stellar age is multiple, separate star formation episodes. Thus, 'Oumuamua may have been close enough to this cloud to have been ejected at the time of one of these episodes given the encounters discussed below.

HD 143978 is one of four YSOs in Lupus we identify as plausible candidates. It moves at 5.34 kms$^{-1}$ (4.25-6.35 kms$^{-1}$ in the 5th to 95th percentiles), and encounters 'Oumuamua 18.87 Myr in the past, at a distance of 11.78 pc (8.66-14.75 in the 5th to 95th percentiles). The disk heating spread in distance is $\sim$ 32-36 pc 18 Myr in the past. HD 143978 is a putative member of the Lupus dark cloud complex, one of the nearest SFRs \citep{galbertei13}. HD 143978's galactic coordinates (l=$339.10\degree$, b=$9.96\degree$) place it in the \textit{on-cloud} component of the Lupus SFR as defined in \cite{galbertei13}, a population of young stars in the immediate vicinity of the Lupus molecular clouds. More specifically, its coordinates indicate that it is a member of the Lupus 3 star forming cloud (see Figures 3 or 15 in \cite{galbertei13}). Although its geometric distance ($\sim$ 97 pc) does not align with the average distance to Lupus 3 (d$_{avg}$=185$^{+11}_{-10}$ pc according to \cite{galbertei13}), \cite{galbertei13} include stars with parallaxes in the range 10-12 mas in defining the membership of Lupus 3 (see their Figure 16). Furthermore, \cite{galbertei13} in their Table 7 exclude it from the population of Lupus stars with doubtful membership status (in Lupus or Upper Centaurus-Lupus) in the literature. We additionally note that its updated parallax measurement from Gaia DR2 places it closer to the Lupus clouds (whose average distances range from $\sim$ 140-200 pc according to \cite{galbertei13}) than the measurement from TYCHO2 used in \cite{galbertei13} (TYCHO2 records its parallax as 11.6$\pm$2.6 mas, placing it 86 pc away, whereas Gaia DR2 measures 10.28$\pm$0.036 mas, placing it 97 pc away). HD 143978 may be a member of the younger TT population, owing to it possibly belonging to the Lupus 3 filament, which was found to harbour a sizeable number of young T-Tau stars \citep{mak07}.

A weak-line T-Tau star in the Lupus SFR, 2MASS J15480212-4004277, was found to pass within 48.19 pc (24.05-92.47 pc in the 5th and 95th percentiles), at 7.09 kms$^{-1}$(6.14-8.69 kms$^{-1}$ in the 5th and 95th percentiles), 23.24 Myr in the past. Between the various disk heating models, dispersion between 'Oumuamua clones takes on a spread of $\sim$ 42-49 pc 23 Myr ago. These conditions were found using Gaia DR2 astrometry for everything but radial velocity, which was taken from observations recorded in \cite{galbertei13} (Table 4, where $v_{r}=2.07 \pm 0.35$ kms$^{-1}$). Using the Gaia DR2 radial velocity produces an encounter with d$^{med}_{enc}$=70.84 pc (28.26-136.95 pc in the 5th and 95th percentiles), 6.95 kms$^{-1}$ (4.57-13.00 kms$^{-1}$ in the 5th and 95th percentiles), 14.97 Myr ago. The Gaia radial velocity however is highly uncertain ($\sigma_{v}$ = 4.2 kms$^{-1}$), and the radial velocity recorded in \cite{galbertei13} falls within these errors. Based on its parallax, it likely belongs to the \textit{off-cloud} component of the Lupus weak-line T-Tau stars \citep{galbertei13}.

2MASS J16081096-3910459 is another weak-line T-Tau star in the Lupus region (\cite{mak07}, \cite{galbertei13}). This passes within 87.35 pc (77.18-99.87 pc in the 5th and 95th percentiles), at 5.98 kms$^{-1}$ (5.56-6.96 kms$^{-1}$ in the 5th and 95th percentiles), 34.03 Myr ago. Disk heating implies a distance spread of $\sim$ 66-78 pc 34 Myr ago; while this star's median distance falls outside this range, we note that the maximum distance spread given by the 'fixed kick' heating model is $\sim$ 100 pc at 34 Myr (see the 'fixed kick' position dispersion for the 2k=2 solution in Figure  \ref{fi:dispersion}). 2MASS J16081096-3910459 has been listed as one of 19 stars which constitute the Lupus core moving group \citep{galbertei13}. This star currently lies $\sim$ 149 pc away. Its coordinates place it either in the Lupus off-cloud weak-line T-Tau star population, or possibly in Lupus 3.

RX J1531.3-3329 has also been listed as a defining member of the Lupus core moving group \citep{galbertei13}. Its coordinates (l=$337.33\degree$, b=$18.49\degree$) place it possibly in the Lupus 1 dark cloud or in the off-cloud population according to Figures 3 and 15 in \cite{galbertei13}. Its distance to 'Oumuamua minimizes at the maximum simulation time however, and is quite large (d$_{enc}^{med}$ = 91.74 pc at -49.99 Myr). RX J1531.3-3329's slow speed of 5.19 kms$^{-1}$ (5.02-5.40 kms$^{-1}$ in the 5th and 95th percentiles) and that of 2MASS J16081096-3910459 indicates that the Lupus moving group moves relatively slowly with respect to 'Oumuamua. 

While the exact depths and spatial distribution of the Lupus clouds is not fully understood \citep{galbertei13}, given the large depth of the Lupus SFR (individual distances to association members range from $\sim$110-190 pc according to \cite{galbertei13}, while \cite{lomladalv08} independently estimate the depth of Lupus to be $51^{+61}_{-35}$ pc), these encounters suggest 'Oumuamua could have interacted with the Lupus 3 or 1 clouds, or perhaps the off-cloud population in the immediate vicinity of the SFR. 

\item{\textbf{Taurus-Auriga SFR}:} 

It has been suggested that Taurus-Auriga has been producing stars for at least 10 Myr \citep{palsta02}. These same authors identify multiple stars in Taurus-Auriga with ages $\sim$ 20 Myr. Some of the older encounters discussed below with Taurus-Auriga stars are therefore too far in the past, but the encounters at 28 Myr and 29 Myr are perhaps not ruled out by these findings. These encounters indicate that 'Oumuamua could have interacted with the Taurus-Auriga molecular clouds, or possibly with the components immediately surrounding the SFR.

V1319 Tau moves at 6.91 kms$^{-1}$ (6.82-7.06 kms$^{-1}$ in the 5th and 95th percentiles), and encounters 'Oumuamua 28.53 Myr in the past, at a distance of 21.79 pc (18.81-25.18 pc in the 5th and 95th percentiles). Its astrometry is taken from its SIMBAD entry, where all measurements but its radial velocity are from Gaia DR2. The SIMBAD record for radial velocity is taken from \cite{ngubraker12}. The Gaia DR2 velocity has much larger uncertainty ($\sigma_{v_{r}}$=6.36 kms$^{-1}$), and produces a much larger spread in encounter conditions. It has an estimated age of 13.00$^{6.50}_{-3.30}$ Myr \citep{davgregre14}, though could be as old as 20 Myr \citep{hamvanpau19}. It has been recorded as a member of the Taurus-Auriga SFR in previous studies \citep{kraherriz17}, but is not included in the recent census provided by \cite{luh18}, possibly owing to its updated Gaia DR2 parallax placing it slightly outside the SFR. Although the canonical distance of the Taurus clouds at $\sim$ 140 pc away \citep{galloiort18} is farther than V1319's geometric distance at $\sim$ 112 pc, the region's depth has been estimated to be at least 20 pc (\cite{kendobhar94}, \cite{torloimio07}, \cite{torloimio12}). Furthermore, \cite{bergen06} found that although the population of classical T Tau stars resides 126-173 pc away, weak line T Tau stars surround the molecular clouds between 106-256 pc. V1319 may therefore reside in the immediate surroundings of the clouds. \cite{kraherriz17} record it as a 'Class III' (disk-free) candidate member of the Taurus-Auriga ecosystem.

2MASS J03190760+3934105 is a T-Tau star, which passes within 16.45 pc (4.85-43.65 pc in the 5th and 95th percentiles), at 6.11 kms $^{-1}$ (5.33-6.61 kms$^{-1}$ in the 5th and 95th percentiles), 29.59 Myr in the past. At $\sim$ 143 pc, it has been suggested to be an outlying member of the Taurus-Auriga SFR, situated on the border of the molecular clouds \citep{lihu98}. Comparing its Right Ascension and Declination to the spatial distribution of Taurus stars provided in \cite{kraherriz17} (Figure 8) indicates that this star is likely situated on the outskirts of the main SFR. However, no newer studies seem to mention a connection between this star and the Taurus SFR, so its classification remains inconclusive.

HD 30171 is a well-known T-Tau star, passing within 46.05 pc (15.96-104.13 pc in the 5th and 95th percentiles), at 8.01 kms$^{-1}$ (6.89-9.39 kms$6{-1}$ in the 5th and 95th percentiles), 43.81 Myr ago. The age of this star has been estimated to be 2-4 Myr \citep{hamvanpau19}, ruling this out as a possible origin.

V1267 Tau is a T-Tau star which has been classified as in the immediate surroundings of the Taurus-Auriga clouds \citep{brojoefer06}. It passes within 40.55 pc (22.89-82.42 pc in the 5th and 95th percentiles), at 7.43 kms$^{-1}$ (6.8-107.83 kms$^{-1}$ in the 5th and 95th percentiles), 49 Myr ago.

\item{\textbf{TW Hya group}:} TYC 8083-45-5 is a bonafide member of the TW Hydra group, and is the only bonafide member of that group we identify in Table \ref{tab:candidate_table}. Of the more than 70 suspected or bonafide members of TW Hydra in our data set, only 3 pass within the distance bounds imposed by this study. This rules out an origin in the TW Hydra group. 

\item{\textbf{Pleiades/Local Association}:} In light of the results of \cite{fenjon18}, it is perhaps unsurprising that 2 encounters at speeds $\sim$ 4-10 kms$^{-1}$ and distances $\sim$ 2-4 pc, and 3 encounters at speeds $\sim$ 10-20 kms$^{-1}$ and distances $\sim$ 4-5 pc, with 5 stars in the Pleiades/Local Association were found; there are 51 stars associated with this group with speeds less than 10 kms$^{-1}$, and 65 with speeds less than 20 kms$^{-1}$ included in the Young Stars Catalog. The dearth of reasonable encounters however point against an origin in this association.

\item{\textbf{Chameleon SFR}:} 

Chamaeleon II has an estimated age of 4$\pm$2 Myr \citep{spealccov08}. The encounter times for the YSOs listed here are much longer than this, indicating that the Cha II region is unlikely to have interacted with 'Oumuamua. 

Sz 46N lies in the Chamaeleon II dark cloud, one of three star-forming clouds in the Chamaeleon system \citep{alcspecha08}. Chamaeleon II lies $178 \pm 18$ pc away \citep{alcspecha08}. Sz 46N encounters 'Oumuamua at a distance of 79.83 pc (48.90-138.60 pc in the 5th to 95th percentiles), 41.89 Myr in the past, moving at 7.02 kms$^{-1}$ (6.49-8.26 kms$^{-1}$ in the 5th and 95th percentiles). We find two other young stars belonging to the Cha II cloud; Hn 23 at 95.08 pc (83.23-106.12 pc in the 5th to 95th percentiles), 45.45 Myr ago, at 6.83 kms$^{-1}$ (6.69-6.99 kms$^{-1}$ in the 5th and 95th percentiles), and CM Cha at 85.38 pc (68.48-127.98 pc in the 5th and 95th percentiles), 44.20 Myr in the past, at 7.50 kms$^{-1}$ (6.95-9.00 kms$^{-1}$ in the 5th and 95th percentiles). 

The age of the Cha I cloud, the most active center of star formation in the Chamaeleon cloud complex encompassing Cha I, II, and III \citep{rei08}, has been estimated to be $\sim$ 2 Myr \citep{rei08}. The Cha I region is, similarly to Cha II, unlikely to be the origin of 'Oumuamua due to its very young age relative to its time of closest approach to 'Oumuamua. 

Hen 3-545, which passes within 69.82 pc, (53.57-81.21 pc in the 5th and 95th percentiles), at 7.40 kms$^{-1}$ (7.27-7.59 kms$^{-1}$ in the 5th and 95th percentiles), 46.81 Myr ago, is a member of the Chamaeleon I star-forming region \citep{mulpasman17}. 2MASS J11054153-7754441 is also a member of the Cha I cloud, passing within 69.37 pc (44.58-91.50 pc in the 5th and 95th percentiles), at 8.17 kms$^{-1}$ (7.89-8.41 kms$^{-1}$ in the 5th and 95th percentiles), 29.75 Myr ago.

In general, the relative speeds of the candidates in the Taurus-Auriga and Chamaeleon SFRs are higher than those from the Lupus SFR.

\item{\textbf{Scorpius Centaurus SFR}:} The age of the subregions of Sco-Cen have been estimated to be 11 $\pm$ 2 Myr for Upper Scorpius, 16 Myr for Upper Centaurus-Lupus, and 17 Myr for Lower Centaurus-Crux \citep{pecmambub12}. CPD-53 5235 is a pre-main sequence star located in the Scorpius-Centaurus region \citep{sonzucbes12}. It passes within 77.29 pc (57.99-102.57 pc in the 5th and 95th percentiles), at 4.32 kms$^{-1}$ (4.09-5.16 kms$^{-1}$ in the 5th and 95th percentiles), 29.41 Myr ago. Given the discrepancy in ages and large distance, we can also rule this out.

\item{\textbf{Other}:} A number of individual stars with no association to moving groups or SFRs were also found. Because many of these have already been considered in previous works, they are not included in Table \ref{tab:candidate_table}. We find very similar candidates from Gaia DR2 as those in \cite{baifarmee18}, and also find similar candidates to those in \cite{zulsansuc18} using SIMBAD data. Here we mention some of the stars which we deem most interesting and have not been thoroughly discussed already in the literature. 

V391 Ori passes within 50.12 pc (43.24-62.10 pc in the 5th and 95th percentiles), 25.52 Myr ago, at 3.84 kms$^{-1}$ (3.77-4.01 kms$^{-1}$ in the 5th and 95th percentiles). The disk heating bounds 26 Myr ago are $\sim$ 49-57 pc. This star's low speed and distance within the disk heating bounds make it a good candidate. V391 Ori has been classified as a weak-line T-Tau star associated with the Orion Nebula Cluster, and was found likely to possess an optically thick accretion disk \citep{szekunrei13}. This star's distance, $\sim$ 100 pc, may indicate V391 Ori is a member of the 32 Orionis moving group; the \texttt{BANYAN} $\Sigma$ moving group analysis code \citep{gagmammal18} gives it a membership probability of 91.6$\%$ (though we note that the \texttt{LACEwING} code \citep{rieblulam18} only gives 4$\%$, predicting a slightly different radial velocity than measured, and placing it 19 pc away from the group's center). The age of 32 Orionis is $\sim$ 15-65 Myr (\cite{rieblulam18} and references therein). Its distance places it some distance away from the nominal value of the Orion Nebula at $\sim$ 400 pc \citep{menreifor07}, so its encounter with 'Oumuamua does not necessarily mean that 'Oumuamua could have interacted with the clouds of Orion. 

BD+11 414 is a T-Tau star passing within 34.61 pc (7.34-89.40 pc in the 5th and 95th percentiles), at 7.73 kms$^{-1}$ (7.09-9.73 kms$^{-1}$ in the 5th and 95th percentiles), 20.29 Myr ago. Disk heating implies a spread in distance of $\sim$ 36-41 pc at this time. Its age has been estimated to be $\sim$ 65 Myr \citep{carboumam09}. The large errors in distance make this a less compelling candidate, but it nevertheless has an appropriate age, along with low speed and median distance.

HD 189210 is a young sun-like star, observed by the Kepler mission \citep{frofracat12}. Its age has been estimated to be 100-200 Myr, though the authors note that their analysis could not exclude an age as young as 50 Myr \citep{frofracat12}. It encounters 'Oumuamua at a distance of 82.86 pc (80.41-90.57 pc in the 5th and 95th percentiles), 49.99 Myr ago, moving at an exceptionally low speed of just 1.84 kms$^{-1}$ (1.69-2.51 kms$^{-1}$ in the 5th and 95th percentiles). Despite the very low speed, the large errors due to disk heating 50 Myr in the past mean this star is not likely the home of 'Oumuamua.

HD 24260 is a member of the population of local white dwarfs \citep{guozhatzi15}, passing within 59.28 pc (33.59-106.68 pc in the 5th and 95th percentiles), at 5.09 kms$^{-1}$ (4.42-5.67 kms$^{-1}$ in the 5th and 95th percentiles), 30 Myr ago. Ejection speeds from white dwarf tidal disruption events of interstellar asteroids have been predicted to be $\sim$ 10-30 kms$^{-1}$ \citep{raf18}. Given this range of expected ejection speeds, this is quite a low encounter speed, though the large spread in distance makes this a less compelling encounter.

EGGR 268 is also a member of the population of local white dwarfs (\cite{holoswsio02}, \cite{oswsiomcc}). The encounter occurs at 3.28 pc (2.27-6.73 pc in the 5th and 95th percentiles), 650 kyr ago, at 24.62 kms$^{-1}$ (11.93-37.01 kms$^{-1}$). Its radial velocity is highly uncertain. It is also a binary system with an M dwarf companion \citep{oswsiomcc}. The spread in encounter distance so early in the past rules this out as an origin.

\end{itemize}

\subsubsection{The Origin of 2I/Borisov}

Using the same code and data sets we have also probed the origin of the second interstellar object discovered, comet 2I/Borisov \citep{guzdrarus19}. The orbital elements and covariance matrix of the comet were obtained from JPL on 2 October 2019 and back-integrated in the manner described in section~\ref{initialconditions}. 

\begin{figure}[ht!]
\plottwo{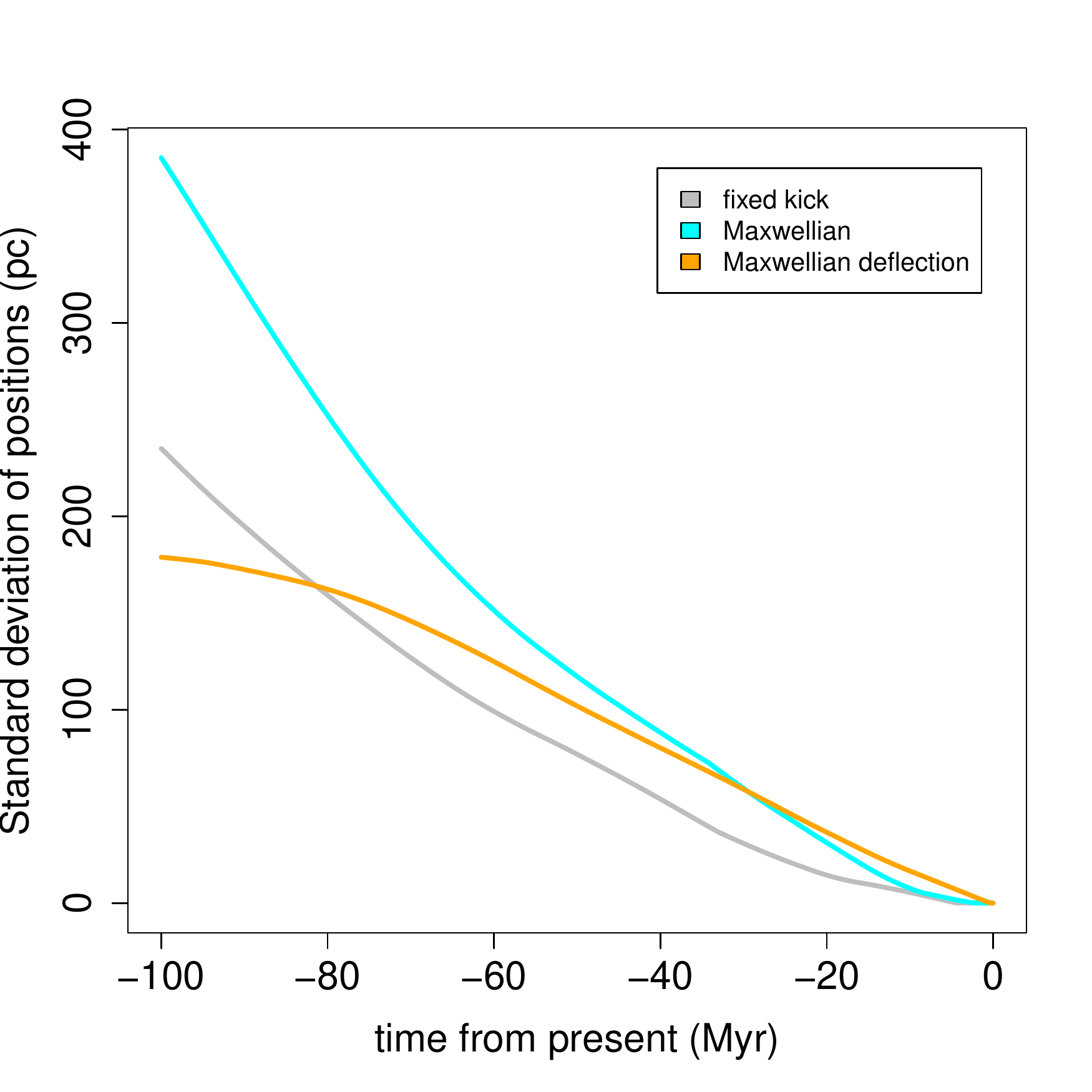}{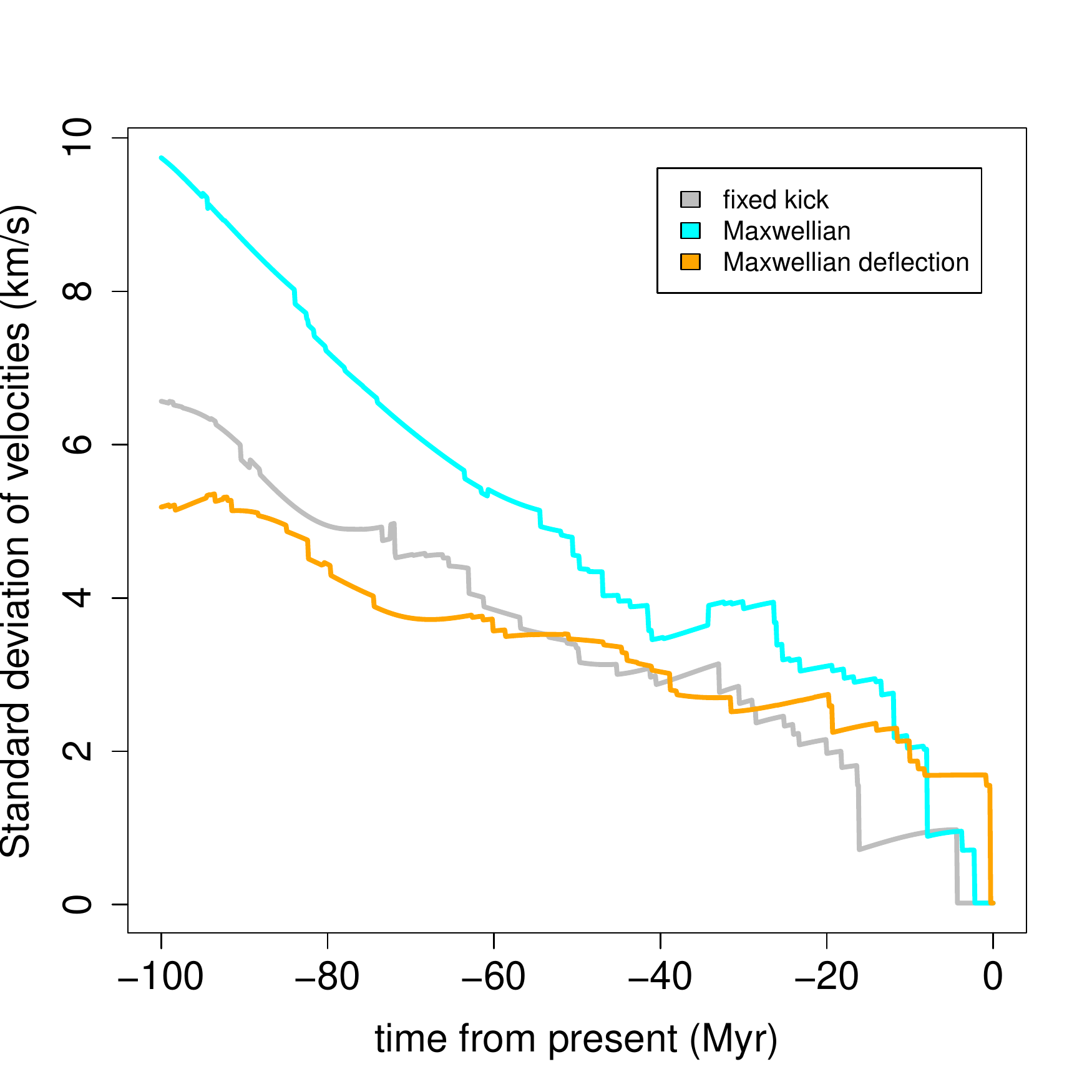}
\caption{The standard deviation of the position and velocity of clones of 2I/Borisov integrated backwards within the Galaxy. \label{fi:2I_dispersion}}
\end{figure}

The dispersion of 100 clones of this comet is presented in Figure~\ref{fi:2I_dispersion}. The behaviour is similar to that of 'Oumuamua, with its dispersion in both position and velocity increasing at a similar rate, despite its larger speed with respect to the LSR. This is to be expected as it suffers the same magnitude impulses at the same rate as 'Oumuamua in our adopted disk heating models.

As for specific potential origin systems, none of the candidates discussed below move at slow enough speed to indicate a likely origin of Borisov. We find the Ursa Major group to have the slowest relative speed. The average encounter distance for the stars within the disk heating bounds is 15.65 pc, 133 kyr ago, at 26.10 kms$^{-1}$. Every other group had average encounter speeds higher than this, typically in the 30-40 kms$^{-1}$ range. Owing to the moving groups' high relative speeds and large distances of encounter, it seems unlikely that this comet comes from any of the nearest young moving groups or kinematic associations. The final stages of its approach to Earth is shown in Animated Figure~\ref{Fig:flythrough_borisov}.

Three stars in the Ursa Major group have plausible encounters with Borisov. GJ 4384 is a multiple star system which passes within 0.32 pc (0.11-0.80 pc in the 5th and 95th percentiles), at 19.15 kms$^{-1}$ (18.74-19.52 kms$^{-1}$ in the 5th and 95th percentiles), 1.46 Myr ago. EV Lac (Gaia DR2 1934263333784036736), a young M dwarf, has a very recent encounter at 0.97 pc (0.96-0.97 pc in the 5th and 95th percentiles), 27.44 kms$^{-1}$ (27.27-27.61 kms$^{-1}$ in the 5th and 95th percentiles), just 150 kyr ago. Lastly, GJ 102, also a flare star, passes within 1.98 pc (1.09-5.27 pc in the 5th and 95th percentiles), at 28.26 kms$^{-1}$ (13.43-44.94 in the 5th and 95th percentiles), 273 kyr ago.

2MASS J03552337+1133437 is a brown dwarf, passing within 0.44 pc (0.32-0.55 pc in the 5th and 95th percentiles), 32.06 kms$^{-1}$ (31.49-32.46 kms$^{-1}$
in the 5th and 95th percentiles), 280 kyr in the past. This is a bonafide member of the AB Dor group \citep{gagmammal18}. 
We find eight stars using Gaia DR2 astrometry which pass within 2 pc at speeds less than 30 kms$^{-1}$ (including EV Lac). Gaia DR2 6223838830917236224 (HD 128356) has the slowest speed at 13.25 kms$^{-1}$, 1.63 pc (1.47-1.79 pc in the 5th and 95th percentiles), 1.95 Myr ago. This is a main sequence star of mass 0.65 M$_{\odot}$ harbouring a Jupiter-mass exoplanet \citep{jenjontuo17}. Given its low mass it likely has a tidal radius smaller than our own ($<$ 1 pc), and considering its relative speed, it is unlikely to be the home of Borisov. There are nine additional Gaia stars which pass within 2 pc at speeds between 30 and 40 kms$^{-1}$. The Gaia stars were only integrated backwards 10 Myr.

These candidates have been summarized Table \ref{tab:borisov_table}, organized according to distance. There are more stars passing within 2 pc at speeds greater than 50 kms$^{-1}$, but due to their high speeds they have not been included in Table \ref{tab:borisov_table}. We do not find any YSOs or stars in SFRs.

In a recent study of possible encounters between 2I/Borisov and Gaia DR2 stars, \cite{baifarye19} find a very close encounter with the star Ross 573, at 0.068 pc (0.053-0.091 pc in the 5th and 95th percentiles), 23 kms$^{-1}$, 910 kyr in the past. We also find a close encounter with this star, recorded as Gaia DR2 5162123155863791744 in our Table \ref{tab:borisov_table}, at similar speed and time but a larger distance of 0.64 pc (0.60-0.69 pc in the 5th and 95th percentiles). As \cite{baifarye19} note, this is likely due to our use of the early, gravity-only solution retrieved on October 2 from JPL, while \cite{baifarye19} account for additional forces as well as a longer data arc. Their second closest encounter is with GJ 4384, at very similar conditions to ours. \cite{baifarye19} note however that GJ 4384 has a physical binary partner, and computing the evolution of the system's center of mass reveals a much larger encounter distance with Borisov than the nominal path of GJ 4384. Our closest encounter, with G 7-34, was found by \cite{baifarye19} to pass at a larger distance of 0.459 pc (0.377-0.580 pc in the 5th and 95th percentiles), with only a 1 percent chance of passing within 0.35 pc. This is slightly different than our encounter at 0.21 pc (0.08-0.32 pc in the 5th and 95th percentiles). An encounter with 2MASS J03552337+1133437 was also found by \cite{baifarye19} at a larger distance of 1.07 pc, differing somewhat from ours at 0.44 pc.  Overall our results are consistent with those of \cite{baifarye19}.

\begin{deluxetable*}{cCCCCCCCCCCCC}
\tablecaption{Our solution for the initial conditions of Borisov in the ICRS frame, at -10 000 years. Coordinates are recorded in Right Ascension ($\alpha$), Declination ($\delta$), distance ($d$), radial velocity ($v_{r}$), and the proper motions in Right Ascension and Declination are given as ($\mu_{\alpha}$, corrected for $\cos\delta$) and ($\mu_{\delta}$), respectively. Each coordinate's 1-$\sigma$ uncertainties are provided.\label{tab:initial_conditions_borisov}}
\tablenum{5}
\tablewidth{0pt}
\tablehead{
\colhead{$\alpha$} & \colhead{$\sigma_{\alpha}$} &
\colhead{$\delta$} & \colhead{$\sigma_{\delta}$} & 
\colhead{$d$} & \colhead{$\sigma_{d}$} & 
\colhead{$v_r$} & \colhead{$\sigma_{v_r}$} & 
\colhead{$\mu_{\alpha}$} & \colhead{$\sigma_{\mu_{\alpha}}$} & 
\colhead{$\mu_{\delta}$} & \colhead{$\sigma_{\mu_{\delta}}$} \\
\colhead{[deg]} & \colhead{[deg]} & 
\colhead{[deg]} & \colhead{[deg]} &
\colhead{[AU]} & \colhead{[AU]} &
\colhead{[kms$^{-1}$]} & \colhead{[kms$^{-1}$]} &
\colhead{[masyr$^{-1}$]} & \colhead{[masyr$^{-1}$]} &
\colhead{[masyr$^{-1}$]} & \colhead{[masyr$^{-1}$]}
}
\startdata
32.6056 & 0.0506 & 59.4884 & 0.0105 & 67646.9119 & 111.2456 & -32.0635 & 0.05274 & 0.8014 & 0.001152 & 0.2305 & 0.0007192 \\
\enddata
\end{deluxetable*}

\begin{longrotatetable}
\begin{deluxetable*}{lllrrrrrrll}
\tablecaption{Plausible local galactic candidates of origin for 'Oumuamua. Plausible systems of origin are listed along with the results of dynamical integrations for the distributions in relative speed and distance. Median times of encounter are also recorded. Along with main identifier, each system has its SIMBAD object type recorded. Membership of systems in star-forming regions (SFRs) or moving groups is also indicated. Bonafide members of moving groups have $^{*}$ beside their identifier; suspected members do not. Stars with (?) included in their \textit{Region/Group} entries have ambiguous membership. These results were obtained using the 2k=2 solution for 'Oumuamua as provided in \cite{baifarmee18}. Astrometric data was taken from \citep{rieblulamcatalog18} and SIMBAD. The results do not change appreciably using our own solution. \label{tab:candidate_table}}
\tablewidth{700pt}
\tabletypesize{\scriptsize}
\tablehead{
\colhead{Identifier} & \colhead{Object Type}& \colhead{$d^{med}_{enc}$} & 
\colhead{$d^{5\%}_{enc}$} & \colhead{$d^{95\%}_{enc}$} & 
\colhead{$v^{med}_{enc}$} & \colhead{$v^{5\%}_{enc}$} & 
\colhead{$v^{95\%}_{enc}$} & \colhead{$t_{enc}$} & 
\colhead{Region/Group} \\ 
\colhead{} & \colhead{} & \colhead{(pc)} & \colhead{(pc)} & \colhead{(pc)} & \colhead{(km s$^{-1}$)} & \colhead{(km s$^{-1}$)} & \colhead{(km s$^{-1}$)} & \colhead{(Myr)} & \colhead{}}
\startdata
HD 143978 & YSO & 11.78 & 8.66 & 14.75 & 5.34 & 4.25 & 6.34 & -18.37 & Lupus SFR \\
2MASS J15480212-4004277 & YSO & 48.59 & 24.05 & 92.47 & 7.09 & 6.14 & 8.69 & -23.24 & Lupus SFR \\
2MASS J16081096-3910459 & T-Tau star & 87.35 & 77.18 & 99.87 & 5.98 & 5.56 & 6.96 & -34.03 & Lupus SFR/moving group \\
RX J1531.3-3329 & T-Tau star & 91.74 & 76.20 & 109.80 & 5.19 & 5.02 & 5.40 & -49.99 & Lupus SFR/moving group \\
V1319 Tau & T-Tau star & 21.79 & 18.81 & 25.18 & 6.91 & 6.82 & 7.06 & -28.53 & Taurus-Auriga SFR \\
1RXS J031907.4+393418 & T-Tau star & 16.45 & 4.85 & 43.65 & 6.11 & 5.33 & 6.61 & -29.59 & Taurus-Auriga SFR(?) \\
HD 30171 & T-Tau star & 46.05 & 15.96 & 104.13 & 8.01 & 6.89 & 9.39 & -43.81 & Taurus-Auriga SFR \\
V1267 Tau & T-Tau star & 40.55 & 22.89 & 82.41 & 7.43 & 6.82 & 7.83 & -49.99 & Taurus-Auriga SFR \\
CPD-53  5235 & pre-MS star & 77.29 & 57.99 & 102.57 & 4.32 & 4.09 & 5.16 & -29.41 & Sco-Cen SFR \\
Hn 23 & YSO & 95.08 & 83.23 & 106.12 & 6.83 & 6.69 & 6.99 & -45.45 & Chamaeleon II SFR \\
Sz 46N & YSO & 79.83 & 48.90 & 138.60 & 7.02 & 6.49 & 8.26 & -41.89 & Chamaeleon II SFR \\
CM Cha & T-Tau star & 85.38 & 68.48 & 127.98 & 7.50 & 6.95 & 9.00 & -44.20 & Chamaeleon II SFR \\
Hen 3-545 & T-Tau star & 69.82 & 53.57 & 81.21 & 7.40 & 7.27 & 7.59 & -46.81 & Chamaeleon I SFR \\
2MASS J11054153-7754441 & Emission-line star & 69.37 & 44.58 & 91.50 & 8.17 & 7.89 & 8.41 & -29.75 & Chamaeleon I SFR \\
HD 49855$^{*}$ & rotationally variable star & 29.81 & 22.12 & 37.38 & 3.57 & 3.48 & 3.65 & -36.50 & Carina  \\
CD-54 4320 & high-pm star & 33.46 & 10.38 & 77.31 & 3.98 & 3.16 & 4.95 & -36.50 & Carina  \\
HD 55279$^{*}$ & rotationally variable star & 44.62 & 36.80 & 55.37 & 4.23 & 4.09 & 4.38 & -27.52 & Carina  \\
V479 Car$^{*}$ & BY Dra variable & 71.66 & 56.89 & 89.32  & 3.43 & 3.32 & 3.72 & -31.90 & Carina  \\
CD-57 1709 & rotationally variable star & 62.95 & 49.65 & 74.38 & 4.90 & 4.34 & 5.63 & -32.96 & Carina  \\
GJ 1167 A & m dwarf & 3.94 & 2.41 & 11.16 & 4.90 & 2.00 & 8.06 & -2.37 & Carina  \\
HD 42270 & rotationally variable star & 49.75 & 37.02 & 65.69 & 4.69 & 4.47 & 4.99 & -26.79 & Carina  \\
HD 83096$^{*}$ & multiple star & 53.64 & 25.84 & 90.36 & 3.57 & 3.26 & 4.18 & -31.02 & Carina  \\
2MASS J09303148-7041479 & rotationally variable star & 61.85 & 39.23 & 85.88 & 5.20 & 4.66 & 5.81 & -29.61 & Carina-vela  \\
CD-53 2515 & rotationally variable star & 81.12 & 24.83 & 130.13 & 6.16 & 4.38 & 7.73 & -34.31 & Carina  \\
HD 83096b$^{*}$ & multiple star & 44.75 & 14.19 & 79.65 & 4.95 & 4.46 & 5.75 & -31.86 & Carina  \\
CD-54 2644 & variable star & 59.65 & 21.37 & 141.31 & 3.52 & 2.37 & 4.65 & -49.14 & Carina  \\
CD-49 4008 & eclipsing binary star & 79.91 & 63.90 & 114.89 & 3.96 & 2.71 & 5.12 & -38.23 & Carina  \\
NOMAD 0269-0122275 & eclipsing binary star & 66.60 & 26.68 & 128.51 & 5.35 & 4.96 & 6.20  & -32.89 & Carina  \\
NOMAD 0331-0114466 & eclipsing binary star & 53.03 & 38.14 & 69.83 & 4.82 & 4.74 & 4.95 & -33.87 & Carina  \\
CD-63 408 & eclipsing binary star & 42.60 & 40.47 & 45.43 & 5.64 & 5.26 & 5.86 & -27.69 & Carina-Vela  \\
HIP 1134$^{*}$ & multiple star & 19.77 & 15.11 & 24.88 & 3.92 & 3.81 & 4.11 & -14.49 & Columba \\
HD 40216$^{*}$ & star & 41.58 & 23.78 & 53.46 & 2.50 & 2.78 & 1.93 & -26.45 & Columba \\
HD 38206$^{*}$ & star & 51.73 & 46.72 & 61.92 & 3.36 & 3.13 & 3.71 & -27.50 & Columba \\
HD 32309$^{*}$ & star & 39.89 & 27.42 & 61.44 & 3.60 & 3.18 & 4.94 & -18.79 & Columba \\
HD 37484$^{*}$ & star & 36.44 & 29.28 & 48.90 & 2.72 & 2.68 & 2.94 & -24.82 & Columba \\
HD 30447$^{*}$ & star & 35.48 & 33.28 & 39.67 & 5.09 & 5.02 & 5.19 & -22.88 & Columba \\
HR 8799(b,c,d,e)$^{*}$ & ellipsoidal variable star $\&$ planets & 25.55 & 16.71 & 51.74 & 2.81 & 2.41 & 4.05 & -18.91 & Columba \\
HD 48370 & high-pm star & 12.92 & 3.51 & 21.28 & 3.46 & 3.25 & 3.64 & -35.85 & Columba \\
HD 37402 & star & 11.91 & 3.46 & 26.95 & 4.84 & 4.30 & 5.39 & -24.46 & Columba \\
HD 32372$^{*}$ & star & 46.24 & 34.25 & 75.64 & 4.96 & 4.43 & 5.84 & -26.56 & Columba \\
kap And A & star & 33.09 & 21.03 & 44.45 & 3.34 & 2.81 & 3.92 & -21.25 & Columba \\
HD 27679 & variable star & 41.49 & 31.81 & 57.46 & 5.78 & 5.29 & 6.22 & -23.42 & Columba \\
HD 36329(a,b) & variable binary stars & 52.49 & 49.80 & 55.04 & 3.90 & 3.70 & 4.07 & -26.41 & Columba \\
CD-48 2324 & rotationally variable star & 71.30 & 52.14 & 154.17 & 3.91 & 3.33 & 4.85 & -49.99 & Columba \\
RBS 595 & rotationally variable star & 46.95 & 39.40 & 56.70 & 5.14 & 4.83 & 5.45 & -25.63 & Columba \\
HD 272836 & rotationally variable star & 44.67 & 23.01 & 84.88 & 4.59 & 4.06 & 5.93 & -25.31 & Columba \\
HD 51797 & rotationally variable star & 66.42 & 57.30 & 84.48 & 2.32 & 2.14 & 2.61 & -49.99 & Columba \\
HD 39130 & star & 61.97 & 33.00 & 91.33 & 5.15 & 4.36 & 6.16 & -29.06 & Columba \\
2MASS J05184616-2756457$^{*}$ & brown dwarf & 50.40 & 36.22 & 58.38 & 3.36 & 1.95 & 4.02 & -24.98 & Columba \\
CD-29 2531 & rotationally variable star & 63.84 & 42.44 & 88.41 & 5.09 & 4.45 & 5.81 & -25.02 & Columba \\
TYC 8157-91-1 & star & 66.54 & 50.89 & 85.93 & 4.48 & 3.48 & 5.48 & -37.42 & Columba \\
TWA 6 & T-Tau star & 26.65 & 20.62 & 63.84 & 3.09 & 2.36 &  5.11 & -20.50 & TW Hydra  \\
eps Scl A/B & high-pm star & 42.94 & 14.21 & 91.39 & 3.60 & 1.45 & 7.53 & -12.33 & TW Hydra  \\
HR 692 & high-pm star & 76.53 & 24.61 & 176.66 & 3.45 & 2.20 & 7.03 & -17.76 & TW Hydra  \\
TYC 8083-45-5 & rotationally variable star & 28.15 & 16.18 & 41.75 & 8.33 & 7.96 & 8.63 & -32.13 & TW Hydra  \\
* b Pup & spectroscopic binary & 21.35 & 6.89 & 163.93 & 7.35 & 3.23 & 15.34 & -17.68 & no group  \\
KIC 4158372 & eruptive variable star & 91.66 & 86.27 & 95.78 & 3.61 & 3.48 & 3.67 & -48.23 & no group  \\
KIC 4929016 & eruptive variable star & 44.62 & 39.74 & 48.85 & 7.03 & 6.83 & 7.17 & -28.64 & no group  \\
HD 189210 & eruptive variable star & 82.86 & 80.41 & 90.57 & 1.84 & 1.69 & 2.51 & -49.99 & no group  \\
2MASS J04405340+2055471 & pre-main sequence star & 58.31 & 41.81 & 91.40 & 4.07 & 3.27 & 5.73 & -20.53 & no group  \\
CVSO 229 & T-Tau star & 84.39 & 65.27 & 29.73 & 16.97 & 5.39 &  4.56 & -49.99 & no group  \\
BD+11 414 & T-Tau star & 34.61 & 7.34 & 89.40 & 7.73 & 7.09 & 9.73 & -20.29 & no group  \\
2MASS J03581272+0932223 & T-Tau star & 65.00 & 36.08 & 147.59 & 6.90 & 3.65 & 9.15 & -49.65 & no group  \\
V391 Ori & T-Tau candidate star & 50.12 & 43.24 & 62.10 & 3.84 & 3.77 & 4.01 & -25.52 & 32 Orionis(?) \\
CD-49  4008 & eclipsing binary & 70.49 & 60.24 & 84.63 & 3.20 & 3.12 & 3.30 & -41.89 & no group  \\
2MASS J00374306-5846229 & brown dwarf & 27.99 & 9.74 & 54.25 & 8.36 & 7.38 & 9.20 & -30.93 & no group  \\
2MASSI J1411213-211950 & low-mass ($< 1 M_{sol}$) star & 18.79 & 5.67 & 45.69 & 3.45 & 1.99 & 7.11 & -11.00 & no group  \\
2MASS J03550477-1032415 & low-mass ($< 1 M_{sol}$) star & 39.68 & 25.18 & 52.78 & 5.09 & 4.42 & 5.67 & -27.40 & no group  \\
HD 24260 & white dwarf & 59.28 & 33.59 & 106.68 & 7.02  & 4.46 & 10.54 & -30.00 & no group  \\
EGGR 268 & white dwarf & 3.28 & 2.27 & 6.73 & 24.62 & 11.93 & 37.01 & -0.65 & no group  \\
\enddata
\end{deluxetable*}
\end{longrotatetable}

\begin{longrotatetable}
\begin{deluxetable*}{lllrrrrrrl}
\tablecaption{Candidate systems of origin for 21/Borisov, organized according to distance. We take all six astrometric coordinates from Gaia DR2 where possible; for G 7-34, GJ 4384, 2MASS J03552337+1133437, and GJ 102, we supplement their Gaia coordinates with radial velocities recorded in SIMBAD. We also supply each SIMBAD object type. These constitute all stars which we find pass within 2 pc at speeds less than 40 kms$^{-1}$. GJ 4384, EV Lac (Gaia DR2 1934263333784036736), and GJ 102 are purported members of the Ursa Major group, and 2MASS J03552337+1133437 is a bonafide member of the AB Dor group. \label{tab:borisov_table}}
\tablewidth{700pt}
\tabletypesize{\scriptsize}
\tablehead{
\colhead{Identifier} & \colhead{Object Type}& \colhead{$d^{med}_{enc}$} & 
\colhead{$d^{5\%}_{enc}$} & \colhead{$d^{95\%}_{enc}$} & 
\colhead{$v^{med}_{enc}$} & \colhead{$v^{5\%}_{enc}$} & 
\colhead{$v^{95\%}_{enc}$} & \colhead{$t_{enc}$} \\ 
\colhead{} & \colhead{} & \colhead{(pc)} & \colhead{(pc)} & \colhead{(pc)} & \colhead{(km s$^{-1}$)} & \colhead{(km s$^{-1}$)} & \colhead{(km s$^{-1}$)} & \colhead{(Myr)}}
\startdata
G 7-34 & flare star & 0.21 & 0.08 & 0.32 & 32.61 & 25.96 & 38.69 & -0.44 \\
GJ 4384 & multiple star & 0.32 & 0.11 & 0.80 & 19.15 & 18.74 & 19.52 & -1.46 \\
2MASS J03552337+1133437 & brown dwarf & 0.44 & 0.32 & 0.55 & 32.06 & 31.49 & 32.46 & -0.28 \\
5162123155863791744 & star & 0.64 & 0.60 & 0.69 & 22.50 & 21.80 & 23.09 & -0.92 \\
411381695718394880 & star & 0.70 & 0.65 & 0.77 & 29.20 & 28.75 & 29.59 & -1.36 \\
3338543951096093696 & high-pm star & 0.72 & 0.34 & 1.33 & 36.75 & 36.25 & 37.39 & -2.90 \\
2176428914377155200 & star & 0.81 & 0.36 & 1.30 & 33.28 & 33.02 & 33.58 & -2.81 \\
207166446152692736 & high-pm star & 0.85 & 0.79 & 0.90 & 35.15 & 34.75 & 35.51 & -1.28 \\
1934263333784036736 & M dwarf/flare star & 0.97 & 0.96 & 0.97 & 27.44 & 27.27 & 27.61 & -0.15 \\
4293318823182081408 & BY Dra variable star & 1.00 & 0.99 & 1.02 & 35.54 & 35.27 & 35.77 & -0.14 \\
978301126629450368 & star & 1.16 & 0.60 & 2.15 & 24.89 & 24.68 & 25.13 & -5.34 \\
350791183318279552 & star & 1.26 & 0.60 & 2.54 & 31.64 & 30.48 & 32.83 & -3.53 \\
217334764042444288 & high-pm star & 1.44 & 1.34 & 1.54 & 35.61 & 35.23 & 36.06 & -0.89 \\
5443030196164951168 & multiple star & 1.54 & 1.41 & 1.72 & 35.18 & 31.21 & 39.06 & -0.45 \\
2612004014832929536 & high-pm star & 1.59 & 0.77 & 2.53 & 38.58 & 38.25 & 38.93 & -3.15 \\
6223838830917236224 & high-pm star & 1.63 & 1.47 & 1.79 & 13.25 & 12.99 & 13.58 & -1.95 \\
461259662021372160 & star & 1.80 & 1.37 & 2.37 & 19.48 & 16.69 & 22.16 & -6.62 \\
4299703274841464320 & high-pm star & 1.83 & 1.38 & 2.28 & 20.02 & 19.42 & 20.67 & -3.15 \\
6501580720836818944 & Multiple star & 1.92 & 1.80 & 2.03 & 25.09 & 24.80 & 25.35 & -1.42 \\
GJ 102 & M dwarf/flare star & 1.98 & 1.09 & 5.27 & 28.26 & 13.43 & 44.94 & -0.27 \\
6249614407131243648 & multiple star & 2.00 & 1.70 & 2.32 & 24.45 & 24.12 & 24.78 & -2.18 \\
\enddata
\end{deluxetable*}
\end{longrotatetable}

\begin{figure*}
    \centering
    \includegraphics[width=8cm]{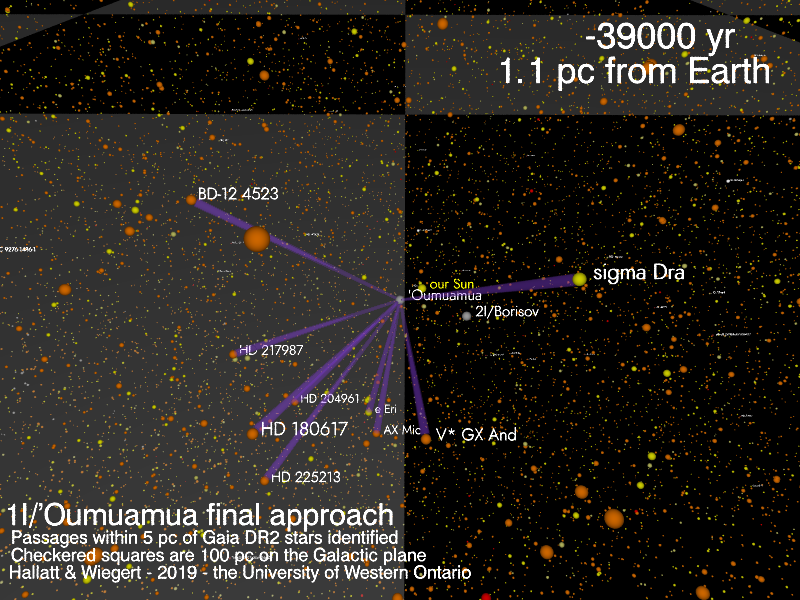}
    \caption{(Animated figure) The final approach of 1I/'Oumuamua to our Solar System. Passages within 5 pc of members of the Gaia DR2 catalog are identified: note that some local stars (e.g. alpha Centauri) are not in the catalog and not shown. The checkerboard pattern is 100 pc on a side and delineates the Galactic plane. In the online version of this animated figure, the animations illustrate the approach of 'Oumuamua to the solar system in 'Oumuamua's frame; close stellar encounters are shown as the ISO and stars move past each other through the galaxy from -300 000 yr to the present, when the Sun is coincident with 'Oumuamua. [This figure is hosted permanently at the Astronomical Journal and semi-permanently at \url{http://www.astro.uwo.ca/~wiegert/interstellar/Gaia-Oumuamua-09.mp4}. Higher resolution versions are available at \url{http://www.astro.uwo.ca/~wiegert/interstellar/}  ] }
    \label{Fig:flythrough_oumuamua}
\end{figure*}

\begin{figure*}
    \centering
    \includegraphics[width=8cm]{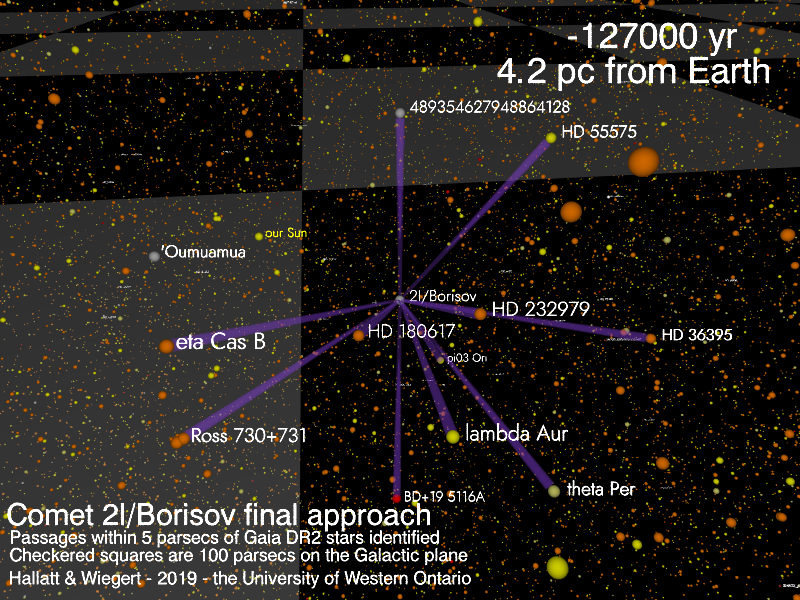}
    \caption{(Animated figure) The final approach of 2I/Borisov to our Solar System, similar to Figure \ref{Fig:flythrough_oumuamua} above for 'Oumuamua. [This figure is hosted permanently at the Astronomical Journal and semi-permanently at \url{http://www.astro.uwo.ca/~wiegert/interstellar/Gaia-Borisov-09.mp4}. Higher resolution versions are available at \url{http://www.astro.uwo.ca/~wiegert/interstellar/}  ] }
    \label{Fig:flythrough_borisov}
\end{figure*}

\section{Conclusions} \label{conclusions}

This paper examines the conjecture that 'Oumuamua is much younger than the age of our Galaxy and that it was ejected at low relative velocity with respect to its origin system. The first part of the conjecture  is based on the asteroid's low speed with respect to the LSR ($< 10$~kms$^{-1}$). Such low velocities are indicative of young $\las 100$~Myr stellar ages from standard age-velocity relations (e.g. \cite{holnorand09,robbiefer17}). The second part of the conjecture, that 'Oumuamua was ejected at low velocity relative to its parent system, is based on the higher efficiency of most ejection processes at launching material at lower speeds.

Given these assumptions, we first modelled the effects of 'disk heating' (gravitational scattering, mostly by giant molecular clouds) on 'Oumuamua's past trajectory to assess our ability to trace it back to its origin, under these stochastic effects. A backwards integration of 'Oumuamua can only expect to be within 15 pc and 2 kms$^{-1}$ of its actual location at -10 Myr, 100~pc and 5 kms$^{-1}$ at -50 Myr, and 400 pc and 10 kms$^{-1}$ at -100 Myr. This limits our ability to retrace its path, but provides a measure of the size of the envelope of potential origin candidates as we go back in time. If 'Oumuamua was indeed ejected at low speed, that means that the path of its origin system parallels that of 'Oumuamua through the galaxy. Though the origin system's trajectory is subject to disk heating as well, this system is currently expected to be within 1 kpc of Earth, within the local Orion Arm, and thus accessible in principle to Earth-based telescopes. Thus we can hope to learn more about this unusual asteroid from telescopic study of its parent system, if such a system can be identified. 

To assess possible local galactic candidate regions, a backwards integration of 'Oumuamua's trajectory was performed together with systems listed in the Gaia DR2 catalog, SIMBAD, as well as the Catalog of Suspected Nearby Young Stars \citep{rieblulamcatalog18} and members of nearby moving groups compiled in \cite{gagmammal18} . Our best candidates are the Carina and Columba moving groups, including the M dwarf GJ 1167 A, the Lupus SFR, and the T-Tau star V391 Ori. 'Oumuamua passes through a considerable subset of the Carina and Columba moving groups at a time comparable to their ages, making them particularly interesting candidate source regions if 'Oumuamua were ejected during planet formation or via intra-cluster interaction in this early stage of the group's history. The Lupus SFR boasts 4 plausible encounters, 2 of which are with weak-line T-Tau stars, which have been found to be part of an older population of the SFR. This older population may hint to multiple episodes of star formation in the clouds \citep{mak07} at roughly the time of 'Oumuamua's encounter, making Lupus another promising candidate.

During the writing of this paper, a second interstellar object, comet 2I/Borisov, was discovered. Though unlikely to be young due to its high velocity with respect to the LSR, back-integrations reveal three stars in the Ursa Major group (GJ 4384, EV Lac, and GJ 102), one brown dwarf of the AB Dor group (2MASS J03552337+113343), and eight Gaia DR2 stars (including EV Lac) to have plausible encounters at speeds $<$ 30 kms$^{-1}$ and within 2 pc. The slowest encounter within 2 pc with Gaia stars is with Gaia DR2 6223838830917236224, at 13 kms$^{-1}$, which is known to harbour a Jupiter-mass exoplanet. Given the high relative speeds of these encounters, it is unlikely any of these stars are the home of 2I/Borisov.

In general, our results for 1I/'Oumuamua are much more consistent with the asteroid being young and local than 2I/Borisov. 'Oumuamua's velocity mirrors many stars' in young, nearby kinematic associations and star forming regions, pointing indirectly to a local origin. As a result, the candidates of origin for 'Oumuamua are more consistent than those for Borisov; the ejection speeds from planetary systems or from intra-cluster interactions are better matched. Future interstellar objects and astrometric surveys will likely provide better opportunity for back-tracking investigations, offering a completely novel means of studying the local galactic neighbourhood.

\acknowledgments

We thank the anonymous referee for a thoughtful and insightful report which much improved this manuscript. This work has made use of data from the European Space Agency (ESA) mission Gaia (https://www.cosmos.esa.int/gaia), processed by the Gaia Data Processing and Analysis Consortium (DPAC, https://www.cosmos.esa.int/web/gaia/dpac/consortium). Funding for the DPAC has been provided by national institutions, in particular the institutions participating in the Gaia Multilateral Agreement. This research has made use of the SIMBAD database, operated at CDS, Strasbourg, France. Funding for this work was provided by the Natural Sciences and Engineering Research Council of Canada (Grant no. RGPIN-2018-05659).

\bibliographystyle{aasjournal}

\bibliography{Wiegert}

\begin{thebibliography}{}
\expandafter\ifx\csname natexlab\endcsname\relax\def\natexlab#1{#1}\fi
\providecommand{\url}[1]{\href{#1}{#1}}

\bibitem[{{Alcal{\'{a}}} {et~al.}(2008){Alcal{\'{a}}}, {Spezzi}, N., {Evans
  II}, {Huard}, K., B., R., E., A., D., \& I.}]{alcspecha08}
{Alcal{\'{a}}}, J., {Spezzi}, L., N., C., {et~al.} 2008, \apj, 676, 427.
\newblock \url{https://doi.org/10.1086$\%$2F527315}

\bibitem[{{Bailer-Jones} {et~al.}(2018){Bailer-Jones}, {Farnocchia}, {Meech},
  {Brasser}, {Micheli}, {Chakrabarti}, {Buie}, \& {Hainaut}}]{baifarmee18}
{Bailer-Jones}, C.~A.~L., {Farnocchia}, D., {Meech}, K.~J., {et~al.} 2018, \aj,
  156, 205

\bibitem[{{Bailer-Jones} {et~al.}(2019){Bailer-Jones}, {Farnocchia}, {Ye},
  {Meech}, \& {Micheli}}]{baifarye19}
{Bailer-Jones}, C. A.~L., {Farnocchia}, D., {Ye}, Q., {Meech}, K.~J., \&
  {Micheli}, M. 2019, arXiv e-prints, arXiv:1912.10213

\bibitem[{{Bertout} \& {Genova}(2006)}]{bergen06}
{Bertout}, C., \& {Genova}, F. 2006, A\&A, 460, 499

\bibitem[{Binney \& Tremaine(1987)}]{bintre87}
Binney, J., \& Tremaine, S. 1987, Galactic Dynamics (Princeton: Princeton
  University Press)

\bibitem[{{Bland-Hawthorn} \& {Gerhard}(2016)}]{blager16}
{Bland-Hawthorn}, J., \& {Gerhard}, O. 2016, Ann. Rev. Astron. Astrophys., 54,
  529

\bibitem[{{Bovy}(2015)}]{bov15}
{Bovy}, J. 2015, Astrophys. J. Suppl., 216, 29

\bibitem[{{Bowler} {et~al.}(2015){Bowler}, {Liu}, {Shkolnik}, \&
  {Tamura}}]{bowliushk15}
{Bowler}, B.~P., {Liu}, M.~C., {Shkolnik}, E.~L., \& {Tamura}, M. 2015, \apjs,
  216, 7

\bibitem[{{Brasser} \& {Morbidelli}(2013)}]{bramor13}
{Brasser}, R., \& {Morbidelli}, A. 2013, Icarus, 225, 40

\bibitem[{{Broeg} {et~al.}(2006){Broeg}, {Joergens}, {Fern{\'a}ndez}, {Husar},
  {Hearty}, {Ammler}, \& {Neuh{\"a}user}}]{brojoefer06}
{Broeg}, C., {Joergens}, V., {Fern{\'a}ndez}, M., {et~al.} 2006, \aap, 450,
  1135

\bibitem[{{Carpenter} {et~al.}(2009){Carpenter}, {Bouwman}, {Mamajek}, {Meyer},
  {Hillenbrand}, {Backman}, {Henning}, {Hines}, {Hollenbach}, {Kim},
  {Moro-Martin}, {Pascucci}, {Silverstone}, {Stauffer}, \&
  {Wolf}}]{carboumam09}
{Carpenter}, J.~M., {Bouwman}, J., {Mamajek}, E.~E., {et~al.} 2009, \apjs, 181,
  197

\bibitem[{{Chen} {et~al.}(2001){Chen}, {Stoughton}, {Smith}, {Uomoto}, {Pier},
  {Yanny}, {Ivezi{\'c}}, {York}, {Anderson}, {Annis}, {Brinkmann}, {Csabai},
  {Fukugita}, {Hindsley}, {Lupton}, {Munn}, \& {SDSS
  Collaboration}}]{chestosmi01}
{Chen}, B., {Stoughton}, C., {Smith}, J.~A., {et~al.} 2001, ApJ, 553, 184

\bibitem[{{Co{\c s}kuno{\v g}lu} {et~al.}(2011){Co{\c s}kuno{\v g}lu}, {Ak},
  {Bilir}, {Karaali}, {Yaz}, {Gilmore}, {Seabroke}, {Bienaym{\'e}},
  {Bland-Hawthorn}, {Campbell}, {Freeman}, {Gibson}, {Grebel}, {Munari},
  {Navarro}, {Parker}, {Siebert}, {Siviero}, {Steinmetz}, {Watson}, {Wyse}, \&
  {Zwitter}}]{cossakbil11}
{Co{\c s}kuno{\v g}lu}, B., {Ak}, S., {Bilir}, S., {et~al.} 2011, \mnras, 412,
  1237

\bibitem[{{Dauphole} \& {Colin}(1995)}]{daucol95}
{Dauphole}, B., \& {Colin}, J. 1995, A\&A, 300, 117

\bibitem[{{Davies} {et~al.}(2014){Davies}, {Gregory}, \&
  {Greaves}}]{davgregre14}
{Davies}, C.~L., {Gregory}, S.~G., \& {Greaves}, J.~S. 2014, MNRAS, 444, 1157

\bibitem[{{Dones} {et~al.}(2004){Dones}, {Weissman}, {Levison}, \&
  {Duncan}}]{donweilev04}
{Dones}, L., {Weissman}, P.~R., {Levison}, H.~F., \& {Duncan}, M.~J. 2004,
  Astronomical Society of the Pacific Conference Series, Vol. 323, {Oort Cloud
  Formation and Dynamics}, ed. D.~{Johnstone}, F.~C. {Adams}, D.~N.~C. {Lin},
  D.~A. {Neufeeld}, \& E.~C. {Ostriker}, 371

\bibitem[{Duncan {et~al.}(1987)Duncan, Quinn, \& Tremaine}]{dunquitre87}
Duncan, M., Quinn, T., \& Tremaine, S. 1987, AJ, 94, 1330

\bibitem[{{Dybczy{\'n}ski} \& {Kr{\'o}likowska}(2018)}]{dybkro18}
{Dybczy{\'n}ski}, P.~A., \& {Kr{\'o}likowska}, M. 2018, \aap, 610, L11

\bibitem[{{Eubanks}(2019)}]{eub19}
{Eubanks}, T.~M. 2019, \apjl, 874, L11

\bibitem[{{Everhart}(1985)}]{eve85}
{Everhart}, E. 1985, in Dynamics of Comets: Their Origin and Evolution, ed.
  A.~{Carusi} \& G.~B. {Valsecchi} (Dordrecht: Kluwer), 185--202

\bibitem[{{Feng} \& {Jones}(2018)}]{fenjon18}
{Feng}, F., \& {Jones}, H.~R.~A. 2018, \apjl, 852, L27

\bibitem[{{Fern{\' a}ndez} \& {Ip}(1984)}]{ferip84}
{Fern{\' a}ndez}, J.~A., \& {Ip}, W.-H. 1984, Icarus, 58, 109

\bibitem[{{Ferri{\`e}re}(2001)}]{fer01}
{Ferri{\`e}re}, K.~M. 2001, Reviews of Modern Physics, 73, 1031

\bibitem[{{Fr{\"o}hlich} {et~al.}(2012){Fr{\"o}hlich}, {Frasca}, {Catanzaro},
  {Bonanno}, {Corsaro}, {Molenda-{\.Z}akowicz}, {Klutsch}, \&
  {Montes}}]{frofracat12}
{Fr{\"o}hlich}, H.-E., {Frasca}, A., {Catanzaro}, G., {et~al.} 2012, A\&A, 543,
  A146

\bibitem[{{Gagn{\'e}} {et~al.}(2018){Gagn{\'e}}, {Mamajek}, {Malo}, {Riedel},
  {Rodriguez}, {Lafreni{\`e}re}, {Faherty}, {Roy-Loubier}, {Pueyo}, {Robin}, \&
  {Doyon}}]{gagmammal18}
{Gagn{\'e}}, J., {Mamajek}, E.~E., {Malo}, L., {et~al.} 2018, \apj, 856, 23

\bibitem[{{Gaia Collaboration} {et~al.}(2016){Gaia Collaboration}, {Prusti},
  {de Bruijne}, {Brown}, {Vallenari}, {Babusiaux}, {Bailer-Jones}, {Bastian},
  {Biermann}, {Evans}, {Eyer}, {Jansen}, {Jordi}, {Klioner}, {Lammers},
  {Lindegren}, {Luri}, {Mignard}, {Milligan}, {Panem}, {Poinsignon},
  {Pourbaix}, {Randich}, {Sarri}, {Sartoretti}, {Siddiqui}, {Soubiran},
  {Valette}, {van Leeuwen}, {Walton}, {Aerts}, {Arenou}, {Cropper}, {Drimmel},
  {H{\o}g}, {Katz}, {Lattanzi}, {O'Mullane}, {Grebel}, {Holland}, {Huc},
  {Passot}, {Bramante}, {Cacciari}, {Casta{\~n}eda}, {Chaoul}, {Cheek}, {De
  Angeli}, {Fabricius}, {Guerra}, {Hern{\'a}ndez}, {Jean-Antoine-Piccolo},
  {Masana}, {Messineo}, {Mowlavi}, {Nienartowicz}, {Ord{\'o}{\~n}ez-Blanco},
  {Panuzzo}, {Portell}, {Richards}, {Riello}, {Seabroke}, {Tanga},
  {Th{\'e}venin}, {Torra}, {Els}, {Gracia-Abril}, {Comoretto},
  {Garcia-Reinaldos}, {Lock}, {Mercier}, {Altmann}, {Andrae}, {Astraatmadja},
  {Bellas-Velidis}, {Benson}, {Berthier}, {Blomme}, {Busso}, {Carry},
  {Cellino}, {Clementini}, {Cowell}, {Creevey}, {Cuypers}, {Davidson}, {De
  Ridder}, {de Torres}, {Delchambre}, {Dell'Oro}, {Ducourant}, {Fr{\'e}mat},
  {Garc{\'\i}a-Torres}, {Gosset}, {Halbwachs}, {Hambly}, {Harrison}, {Hauser},
  {Hestroffer}, {Hodgkin}, {Huckle}, {Hutton}, {Jasniewicz}, {Jordan},
  {Kontizas}, {Korn}, {Lanzafame}, {Manteiga}, {Moitinho}, {Muinonen},
  {Osinde}, {Pancino}, {Pauwels}, {Petit}, {Recio-Blanco}, {Robin}, {Sarro},
  {Siopis}, {Smith}, {Smith}, {Sozzetti}, {Thuillot}, {van Reeven}, {Viala},
  {Abbas}, {Abreu Aramburu}, {Accart}, {Aguado}, {Allan}, {Allasia},
  {Altavilla}, {{\'A}lvarez}, {Alves}, {Anderson}, {Andrei}, {Anglada Varela},
  {Antiche}, {Antoja}, {Ant{\'o}n}, {Arcay}, {Atzei}, {Ayache}, {Bach},
  {Baker}, {Balaguer-N{\'u}{\~n}ez}, {Barache}, {Barata}, {Barbier}, {Barblan},
  {Baroni}, {Barrado y Navascu{\'e}s}, {Barros}, {Barstow}, {Becciani},
  {Bellazzini}, {Bellei}, {Bello Garc{\'\i}a}, {Belokurov}, {Bendjoya},
  {Berihuete}, {Bianchi}, {Bienaym{\'e}}, {Billebaud}, {Blagorodnova},
  {Blanco-Cuaresma}, {Boch}, {Bombrun}, {Borrachero}, {Bouquillon}, {Bourda},
  {Bouy}, {Bragaglia}, {Breddels}, {Brouillet}, {Br{\"u}semeister},
  {Bucciarelli}, {Budnik}, {Burgess}, {Burgon}, {Burlacu}, {Busonero}, {Buzzi},
  {Caffau}, {Cambras}, {Campbell}, {Cancelliere}, {Cantat-Gaudin}, {Carlucci},
  {Carrasco}, {Castellani}, {Charlot}, {Charnas}, {Charvet}, {Chassat},
  {Chiavassa}, {Clotet}, {Cocozza}, {Collins}, {Collins}, {Costigan}, {Crifo},
  {Cross}, {Crosta}, {Crowley}, {Dafonte}, {Damerdji}, {Dapergolas}, {David},
  {David}, {De Cat}, {de Felice}, {de Laverny}, {De Luise}, {De March}, {de
  Martino}, {de Souza}, {Debosscher}, {del Pozo}, {Delbo}, {Delgado},
  {Delgado}, {di Marco}, {Di Matteo}, {Diakite}, {Distefano}, {Dolding}, {Dos
  Anjos}, {Drazinos}, {Dur{\'a}n}, {Dzigan}, {Ecale}, {Edvardsson}, {Enke},
  {Erdmann}, {Escolar}, {Espina}, {Evans}, {Eynard Bontemps}, {Fabre},
  {Fabrizio}, {Faigler}, {Falc{\~a}o}, {Farr{\`a}s Casas}, {Faye}, {Federici},
  {Fedorets}, {Fern{\'a}ndez-Hern{\'a}ndez}, {Fernique}, {Fienga}, {Figueras},
  {Filippi}, {Findeisen}, {Fonti}, {Fouesneau}, {Fraile}, {Fraser}, {Fuchs},
  {Furnell}, {Gai}, {Galleti}, {Galluccio}, {Garabato}, {Garc{\'\i}a-Sedano},
  {Gar{\'e}}, {Garofalo}, {Garralda}, {Gavras}, {Gerssen}, {Geyer}, {Gilmore},
  {Girona}, {Giuffrida}, {Gomes}, {Gonz{\'a}lez-Marcos},
  {Gonz{\'a}lez-N{\'u}{\~n}ez}, {Gonz{\'a}lez-Vidal}, {Granvik}, {Guerrier},
  {Guillout}, {Guiraud}, {G{\'u}rpide}, {Guti{\'e}rrez-S{\'a}nchez}, {Guy},
  {Haigron}, {Hatzidimitriou}, {Haywood}, {Heiter}, {Helmi}, {Hobbs},
  {Hofmann}, {Holl}, {Holland }, {Hunt}, {Hypki}, {Icardi}, {Irwin}, {Jevardat
  de Fombelle}, {Jofr{\'e}}, {Jonker}, {Jorissen}, {Julbe}, {Karampelas},
  {Kochoska}, {Kohley}, {Kolenberg}, {Kontizas}, {Koposov}, {Kordopatis},
  {Koubsky}, {Kowalczyk}, {Krone-Martins}, {Kudryashova}, {Kull}, {Bachchan},
  {Lacoste-Seris}, {Lanza}, {Lavigne}, {Le Poncin-Lafitte}, {Lebreton},
  {Lebzelter}, {Leccia}, {Leclerc}, {Lecoeur-Taibi}, {Lemaitre}, {Lenhardt},
  {Leroux}, {Liao}, {Licata}, {Lindstr{\o}m}, {Lister}, {Livanou}, {Lobel},
  {L{\"o}ffler}, {L{\'o}pez}, {Lopez-Lozano}, {Lorenz}, {Loureiro},
  {MacDonald}, {Magalh{\~a}es Fernandes}, {Managau}, {Mann}, {Mantelet},
  {Marchal}, {Marchant}, {Marconi}, {Marie}, {Marinoni}, {Marrese},
  {Marschalk{\'o}}, {Marshall}, {Mart{\'\i}n-Fleitas}, {Martino}, {Mary},
  {Matijevi{\v{c}}}, {Mazeh}, {McMillan}, {Messina}, {Mestre}, {Michalik},
  {Millar}, {Miranda}, {Molina}, {Molinaro}, {Molinaro}, {Moln{\'a}r},
  {Moniez}, {Montegriffo}, {Monteiro}, {Mor}, {Mora}, {Morbidelli}, {Morel},
  {Morgenthaler}, {Morley}, {Morris}, {Mulone}, {Muraveva}, {Musella},
  {Narbonne}, {Nelemans}, {Nicastro}, {Noval}, {Ord{\'e}novic},
  {Ordieres-Mer{\'e}}, {Osborne}, {Pagani}, {Pagano}, {Pailler}, {Palacin},
  {Palaversa}, {Parsons}, {Paulsen}, {Pecoraro}, {Pedrosa}, {Pentik{\"a}inen},
  {Pereira}, {Pichon}, {Piersimoni}, {Pineau}, {Plachy}, {Plum}, {Poujoulet},
  {Pr{\v{s}}a}, {Pulone}, {Ragaini}, {Rago}, {Rambaux}, {Ramos-Lerate},
  {Ranalli}, {Rauw}, {Read}, {Regibo}, {Renk}, {Reyl{\'e}}, {Ribeiro},
  {Rimoldini}, {Ripepi}, {Riva}, {Rixon}, {Roelens}, {Romero-G{\'o}mez},
  {Rowell}, {Royer}, {Rudolph}, {Ruiz-Dern}, {Sadowski}, {Sagrist{\`a}
  Sell{\'e}s}, {Sahlmann}, {Salgado}, {Salguero}, {Sarasso}, {Savietto},
  {Schnorhk}, {Schultheis}, {Sciacca}, {Segol}, {Segovia}, {Segransan},
  {Serpell}, {Shih}, {Smareglia}, {Smart}, {Smith}, {Solano}, {Solitro},
  {Sordo}, {Soria Nieto}, {Souchay}, {Spagna}, {Spoto}, {Stampa}, {Steele},
  {Steidelm{\"u}ller}, {Stephenson}, {Stoev}, {Suess}, {S{\"u}veges}, {Surdej},
  {Szabados}, {Szegedi-Elek}, {Tapiador}, {Taris}, {Tauran}, {Taylor},
  {Teixeira}, {Terrett}, {Tingley}, {Trager}, {Turon}, {Ulla}, {Utrilla},
  {Valentini}, {van Elteren}, {Van Hemelryck}, {van Leeuwen}, {Varadi},
  {Vecchiato}, {Veljanoski}, {Via}, {Vicente}, {Vogt}, {Voss}, {Votruba},
  {Voutsinas}, {Walmsley}, {Weiler}, {Weingrill}, {Werner}, {Wevers},
  {Whitehead}, {Wyrzykowski}, {Yoldas}, {{\v{Z}}erjal}, {Zucker}, {Zurbach},
  {Zwitter}, {Alecu}, {Allen}, {Allende Prieto}, {Amorim},
  {Anglada-Escud{\'e}}, {Arsenijevic}, {Azaz}, {Balm}, {Beck}, {Bernstein},
  {Bigot}, {Bijaoui}, {Blasco}, {Bonfigli}, {Bono}, {Boudreault}, {Bressan},
  {Brown}, {Brunet}, {Bunclark}, {Buonanno}, {Butkevich}, {Carret}, {Carrion},
  {Chemin}, {Ch{\'e}reau}, {Corcione}, {Darmigny}, {de Boer}, {de Teodoro}, {de
  Zeeuw}, {Delle Luche}, {Domingues}, {Dubath}, {Fodor}, {Fr{\'e}zouls},
  {Fries}, {Fustes}, {Fyfe}, {Gallardo}, {Gallegos}, {Gardiol}, {Gebran},
  {Gomboc}, {G{\'o}mez}, {Grux}, {Gueguen}, {Heyrovsky}, {Hoar}, {Iannicola},
  {Isasi Parache}, {Janotto}, {Joliet}, {Jonckheere}, {Keil}, {Kim},
  {Klagyivik}, {Klar}, {Knude}, {Kochukhov}, {Kolka}, {Kos}, {Kutka}, {Lainey},
  {LeBouquin}, {Liu}, {Loreggia}, {Makarov}, {Marseille}, {Martayan},
  {Martinez-Rubi}, {Massart}, {Meynadier}, {Mignot}, {Munari}, {Nguyen},
  {Nordlander}, {Ocvirk}, {O'Flaherty}, {Olias Sanz}, {Ortiz}, {Osorio},
  {Oszkiewicz}, {Ouzounis}, {Palmer}, {Park}, {Pasquato}, {Peltzer}, {Peralta},
  {P{\'e}turaud}, {Pieniluoma}, {Pigozzi}, {Poels}, {Prat}, {Prod'homme},
  {Raison}, {Rebordao}, {Risquez}, {Rocca-Volmerange}, {Rosen}, {Ruiz-Fuertes},
  {Russo}, {Sembay}, {Serraller Vizcaino}, {Short}, {Siebert}, {Silva},
  {Sinachopoulos}, {Slezak}, {Soffel}, {Sosnowska}, {Strai{\v{z}}ys}, {ter
  Linden}, {Terrell}, {Theil}, {Tiede}, {Troisi}, {Tsalmantza}, {Tur},
  {Vaccari}, {Vachier}, {Valles}, {Van Hamme}, {Veltz}, {Virtanen}, {Wallut},
  {Wichmann}, {Wilkinson}, {Ziaeepour}, \& {Zschocke}}]{Gaia16}
{Gaia Collaboration}, {Prusti}, T., {de Bruijne}, J.~H.~J., {et~al.} 2016,
  \aap, 595, A1

\bibitem[{{Gaia Collaboration} {et~al.}(2018){Gaia Collaboration}, {Brown},
  {Vallenari}, {Prusti}, {de Bruijne}, {Babusiaux}, {Bailer-Jones}, {Biermann},
  {Evans}, {Eyer}, {Jansen}, {Jordi}, {Klioner}, {Lammers}, {Lindegren},
  {Luri}, {Mignard}, {Panem}, {Pourbaix}, {Randich}, {Sartoretti}, {Siddiqui},
  {Soubiran}, {van Leeuwen}, {Walton}, {Arenou}, {Bastian}, {Cropper},
  {Drimmel}, {Katz}, {Lattanzi}, {Bakker}, {Cacciari}, {Casta{\~n}eda},
  {Chaoul}, {Cheek}, {De Angeli}, {Fabricius}, {Guerra}, {Holl}, {Masana},
  {Messineo}, {Mowlavi}, {Nienartowicz}, {Panuzzo}, {Portell}, {Riello},
  {Seabroke}, {Tanga}, {Th{\'e}venin}, {Gracia-Abril}, {Comoretto},
  {Garcia-Reinaldos}, {Teyssier}, {Altmann}, {Andrae}, {Audard},
  {Bellas-Velidis}, {Benson}, {Berthier}, {Blomme}, {Burgess}, {Busso},
  {Carry}, {Cellino}, {Clementini}, {Clotet}, {Creevey}, {Davidson}, {De
  Ridder}, {Delchambre}, {Dell'Oro}, {Ducourant},
  {Fern{\'a}ndez-Hern{\'a}ndez}, {Fouesneau}, {Fr{\'e}mat}, {Galluccio},
  {Garc{\'\i}a-Torres}, {Gonz{\'a}lez-N{\'u}{\~n}ez}, {Gonz{\'a}lez-Vidal},
  {Gosset}, {Guy}, {Halbwachs}, {Hambly}, {Harrison}, {Hern{\'a}ndez},
  {Hestroffer}, {Hodgkin}, {Hutton}, {Jasniewicz}, {Jean-Antoine-Piccolo},
  {Jordan}, {Korn}, {Krone-Martins}, {Lanzafame}, {Lebzelter}, {L{\"o}ffler},
  {Manteiga}, {Marrese}, {Mart{\'\i}n-Fleitas}, {Moitinho}, {Mora}, {Muinonen},
  {Osinde}, {Pancino}, {Pauwels}, {Petit}, {Recio-Blanco}, {Richards},
  {Rimoldini}, {Robin}, {Sarro}, {Siopis}, {Smith}, {Sozzetti}, {S{\"u}veges},
  {Torra}, {van Reeven}, {Abbas}, {Abreu Aramburu}, {Accart}, {Aerts},
  {Altavilla}, {{\'A}lvarez}, {Alvarez}, {Alves}, {Anderson}, {Andrei},
  {Anglada Varela}, {Antiche}, {Antoja}, {Arcay}, {Astraatmadja}, {Bach},
  {Baker}, {Balaguer-N{\'u}{\~n}ez}, {Balm}, {Barache}, {Barata}, {Barbato},
  {Barblan}, {Barklem}, {Barrado}, {Barros}, {Barstow}, {Bartholom{\'e}
  Mu{\~n}oz}, {Bassilana}, {Becciani}, {Bellazzini}, {Berihuete}, {Bertone},
  {Bianchi}, {Bienaym{\'e}}, {Blanco-Cuaresma}, {Boch}, {Boeche}, {Bombrun},
  {Borrachero}, {Bossini}, {Bouquillon}, {Bourda}, {Bragaglia}, {Bramante},
  {Breddels}, {Bressan}, {Brouillet}, {Br{\"u}semeister}, {Brugaletta},
  {Bucciarelli}, {Burlacu}, {Busonero}, {Butkevich}, {Buzzi}, {Caffau},
  {Cancelliere}, {Cannizzaro}, {Cantat-Gaudin}, {Carballo}, {Carlucci},
  {Carrasco}, {Casamiquela}, {Castellani}, {Castro-Ginard}, {Charlot},
  {Chemin}, {Chiavassa}, {Cocozza}, {Costigan}, {Cowell}, {Crifo}, {Crosta},
  {Crowley}, {Cuypers}, {Dafonte}, {Damerdji}, {Dapergolas}, {David}, {David},
  {de Laverny}, {De Luise}, {De March}, {de Martino}, {de Souza}, {de Torres},
  {Debosscher}, {del Pozo}, {Delbo}, {Delgado}, {Delgado}, {Di Matteo},
  {Diakite}, {Diener}, {Distefano}, {Dolding}, {Drazinos}, {Dur{\'a}n},
  {Edvardsson}, {Enke}, {Eriksson}, {Esquej}, {Eynard Bontemps}, {Fabre},
  {Fabrizio}, {Faigler}, {Falc{\~a}o}, {Farr{\`a}s Casas}, {Federici},
  {Fedorets}, {Fernique}, {Figueras}, {Filippi}, {Findeisen}, {Fonti},
  {Fraile}, {Fraser}, {Fr{\'e}zouls}, {Gai}, {Galleti}, {Garabato},
  {Garc{\'\i}a-Sedano}, {Garofalo}, {Garralda}, {Gavel}, {Gavras}, {Gerssen},
  {Geyer}, {Giacobbe}, {Gilmore}, {Girona}, {Giuffrida}, {Glass}, {Gomes},
  {Granvik}, {Gueguen}, {Guerrier}, {Guiraud}, {Guti{\'e}rrez-S{\'a}nchez},
  {Haigron}, {Hatzidimitriou}, {Hauser}, {Haywood}, {Heiter}, {Helmi}, {Heu},
  {Hilger}, {Hobbs}, {Hofmann}, {Holland}, {Huckle}, {Hypki}, {Icardi},
  {Jan{\ss}en}, {Jevardat de Fombelle}, {Jonker}, {Juh{\'a}sz}, {Julbe},
  {Karampelas}, {Kewley}, {Klar}, {Kochoska}, {Kohley}, {Kolenberg},
  {Kontizas}, {Kontizas}, {Koposov}, {Kordopatis}, {Kostrzewa-Rutkowska},
  {Koubsky}, {Lambert}, {Lanza}, {Lasne}, {Lavigne}, {Le Fustec}, {Le
  Poncin-Lafitte}, {Lebreton}, {Leccia}, {Leclerc}, {Lecoeur-Taibi},
  {Lenhardt}, {Leroux}, {Liao}, {Licata}, {Lindstr{\o}m}, {Lister}, {Livanou},
  {Lobel}, {L{\'o}pez}, {Managau}, {Mann}, {Mantelet}, {Marchal}, {Marchant},
  {Marconi}, {Marinoni}, {Marschalk{\'o}}, {Marshall}, {Martino}, {Marton},
  {Mary}, {Massari}, {Matijevi{\v{c}}}, {Mazeh}, {McMillan}, {Messina},
  {Michalik}, {Millar}, {Molina}, {Molinaro}, {Moln{\'a}r}, {Montegriffo},
  {Mor}, {Morbidelli}, {Morel}, {Morris}, {Mulone}, {Muraveva}, {Musella},
  {Nelemans}, {Nicastro}, {Noval}, {O'Mullane}, {Ord{\'e}novic},
  {Ord{\'o}{\~n}ez-Blanco}, {Osborne}, {Pagani}, {Pagano}, {Pailler},
  {Palacin}, {Palaversa}, {Panahi}, {Pawlak}, {Piersimoni}, {Pineau}, {Plachy},
  {Plum}, {Poggio}, {Poujoulet}, {Pr{\v{s}}a}, {Pulone}, {Racero}, {Ragaini},
  {Rambaux}, {Ramos-Lerate}, {Regibo}, {Reyl{\'e}}, {Riclet}, {Ripepi}, {Riva},
  {Rivard}, {Rixon}, {Roegiers}, {Roelens}, {Romero-G{\'o}mez}, {Rowell},
  {Royer}, {Ruiz-Dern}, {Sadowski}, {Sagrist{\`a} Sell{\'e}s}, {Sahlmann},
  {Salgado}, {Salguero}, {Sanna}, {Santana-Ros}, {Sarasso}, {Savietto},
  {Schultheis}, {Sciacca}, {Segol}, {Segovia}, {S{\'e}gransan}, {Shih},
  {Siltala}, {Silva}, {Smart}, {Smith}, {Solano}, {Solitro}, {Sordo}, {Soria
  Nieto}, {Souchay}, {Spagna}, {Spoto}, {Stampa}, {Steele},
  {Steidelm{\"u}ller}, {Stephenson}, {Stoev}, {Suess}, {Surdej}, {Szabados},
  {Szegedi-Elek}, {Tapiador}, {Taris}, {Tauran}, {Taylor}, {Teixeira},
  {Terrett}, {Teyssand ier}, {Thuillot}, {Titarenko}, {Torra Clotet}, {Turon},
  {Ulla}, {Utrilla}, {Uzzi}, {Vaillant}, {Valentini}, {Valette}, {van Elteren},
  {Van Hemelryck}, {van Leeuwen}, {Vaschetto}, {Vecchiato}, {Veljanoski},
  {Viala}, {Vicente}, {Vogt}, {von Essen}, {Voss}, {Votruba}, {Voutsinas},
  {Walmsley}, {Weiler}, {Wertz}, {Wevers}, {Wyrzykowski}, {Yoldas},
  {{\v{Z}}erjal}, {Ziaeepour}, {Zorec}, {Zschocke}, {Zucker}, {Zurbach}, \&
  {Zwitter}}]{gaiaDR2}
{Gaia Collaboration}, {Brown}, A.~G.~A., {Vallenari}, A., {et~al.} 2018, \aap,
  616, A1

\bibitem[{{Gaidos}(2018)}]{gai18}
{Gaidos}, E. 2018, \mnras, 477, 5692

\bibitem[{{Gaidos} {et~al.}(2017){Gaidos}, {Williams}, \&
  {Kraus}}]{gaiwilkra17}
{Gaidos}, E., {Williams}, J., \& {Kraus}, A. 2017, Research Notes of the
  American Astronomical Society, 1, 13

\bibitem[{{Galli} {et~al.}(2013){Galli}, {Bertout}, {Teixeira}, \&
  {Ducourant}}]{galbertei13}
{Galli}, P.~A.~B., {Bertout}, C., {Teixeira}, R., \& {Ducourant}, C. 2013,
  \aap, 558, A77

\bibitem[{{Galli} {et~al.}(2018){Galli}, {Loinard}, {Ortiz-L{\'e}on},
  {Kounkel}, {Dzib}, {Mioduszewski}, {Rodr{\'{\i}}guez}, {Hartmann},
  {Teixeira}, {Torres}, {Rivera}, {Boden}, {Evans}, {Brice{\~n}o}, {Tobin}, \&
  {Heyer}}]{galloiort18}
{Galli}, P.~A.~B., {Loinard}, L., {Ortiz-L{\'e}on}, G.~N., {et~al.} 2018, ApJ,
  859, 33

\bibitem[{{Gammie} {et~al.}(1991){Gammie}, {Ostriker}, \& {Jog}}]{gamostjog91}
{Gammie}, C.~F., {Ostriker}, J.~P., \& {Jog}, C.~J. 1991, \apj, 378, 565

\bibitem[{{Gillessen} {et~al.}(2009){Gillessen}, {Eisenhauer}, {Trippe},
  {Alexander}, {Genzel}, {Martins}, \& {Ott}}]{gileistri09}
{Gillessen}, S., {Eisenhauer}, F., {Trippe}, S., {et~al.} 2009, ApJ, 692, 1075

\bibitem[{{Guo} {et~al.}(2015){Guo}, {Zhao}, {Tziamtzis}, {Liu}, {Li}, {Zhang},
  {Hou}, \& {Wang}}]{guozhatzi15}
{Guo}, J., {Zhao}, J., {Tziamtzis}, A., {et~al.} 2015, \mnras, 454, 2787

\bibitem[{{Guzik} {et~al.}(2019){Guzik}, {Drahus}, {Rusek}, {Waniak},
  {Cannizzaro}, \& {Pastor-Marazuela}}]{guzdrarus19}
{Guzik}, P., {Drahus}, M., {Rusek}, K., {et~al.} 2019, Nature Astronomy,
  doi:10.1038/s41550-019-0931-8

\bibitem[{{Hamb{\'a}lek} {et~al.}(2019){Hamb{\'a}lek}, {Va{\AA}ko}, {Paunzen},
  \& {Smalley}}]{hamvanpau19}
{Hamb{\'a}lek}, {\"A}., {Va{\AA}ko}, M., {Paunzen}, E., \& {Smalley}, B. 2019,
  MNRAS, 483, 1642

\bibitem[{{Hands} {et~al.}(2019){Hands}, {Dehnen}, {Gration}, {Stadel}, \&
  {Moore}}]{handehgra19}
{Hands}, T., {Dehnen}, W., {Gration}, A., {Stadel}, J., \& {Moore}, B. 2019,
  \mnras, 490, 21

\bibitem[{{Holberg} {et~al.}(2002){Holberg}, {Oswalt}, \& {Sion}}]{holoswsio02}
{Holberg}, J.~B., {Oswalt}, T.~D., \& {Sion}, E.~M. 2002, ApJ, 571, 512

\bibitem[{{Holmberg} {et~al.}(2009){Holmberg}, {Nordstr{\"o}m}, \&
  {Andersen}}]{holnorand09}
{Holmberg}, J., {Nordstr{\"o}m}, B., \& {Andersen}, J. 2009, A\&A, 501, 941

\bibitem[{{Hughes} {et~al.}(1994){Hughes}, {Hartigan}, {Krautter}, \&
  {Kelemen}}]{hugharkra94}
{Hughes}, J., {Hartigan}, P., {Krautter}, J., \& {Kelemen}, J. 1994, AJ, 108,
  1071

\bibitem[{{Jackson} {et~al.}(2018){Jackson}, {Tamayo}, {Hammond}, {Ali-Dib}, \&
  {Rein}}]{jactamhamalirei18}
{Jackson}, A., {Tamayo}, D., {Hammond}, N., {Ali-Dib}, M., \& {Rein}, H. 2018,
  MNRAS, 478, L49

\bibitem[{{Jenkins} {et~al.}(2017){Jenkins}, {Jones}, {Tuomi}, {D{\'\i}az},
  {Cordero}, {Aguayo}, {Pantoja}, {Arriagada}, {Mahu}, {Brahm}, {Rojo}, {Soto},
  {Ivanyuk}, {Becerra Yoma}, {Day-Jones}, {Ruiz}, {Pavlenko}, {Barnes},
  {Murgas}, {Pinfield}, {Jones}, {L{\'o}pez-Morales}, {Shectman}, {Butler}, \&
  {Minniti}}]{jenjontuo17}
{Jenkins}, J.~S., {Jones}, H.~R.~A., {Tuomi}, M., {et~al.} 2017, \mnras, 466,
  443

\bibitem[{{Jog} \& {Ostriker}(1988)}]{jogost88}
{Jog}, C.~J., \& {Ostriker}, J.~P. 1988, \apj, 328, 404

\bibitem[{{Kenyon} {et~al.}(1994){Kenyon}, {Dobrzycka}, \&
  {Hartmann}}]{kendobhar94}
{Kenyon}, S.~J., {Dobrzycka}, D., \& {Hartmann}, L. 1994, AJ, 108, 1872

\bibitem[{{Kraus} {et~al.}(2017){Kraus}, {Herczeg}, {Rizzuto}, {Mann},
  {Slesnick}, {Carpenter}, {Hillenbrand}, \& {Mamajek}}]{kraherriz17}
{Kraus}, A.~L., {Herczeg}, G.~J., {Rizzuto}, A.~C., {et~al.} 2017, ApJ, 838,
  150

\bibitem[{{Li} \& {Hu}(1998)}]{lihu98}
{Li}, J.~Z., \& {Hu}, J.~Y. 1998, Astron. Astrophys. Suppl., 132, 173

\bibitem[{{Lombardi} {et~al.}(2008){Lombardi}, {Lada}, \&
  {Alves}}]{lomladalv08}
{Lombardi}, M., {Lada}, C.~J., \& {Alves}, J. 2008, A\&A, 480, 785

\bibitem[{{Luhman}(2018)}]{luh18}
{Luhman}, K. 2018, \aj, 156, 271

\bibitem[{{Makarov}(2007)}]{mak07}
{Makarov}, V. 2007, \apj, 658, 480

\bibitem[{{Mamajek}(2017)}]{mam17}
{Mamajek}, E. 2017, Research Notes of the American Astronomical Society, 1, 21

\bibitem[{{Meech} {et~al.}(2017){Meech}, {Weryk}, {Micheli}, {Kleyna},
  {Hainaut}, {Jedicke}, {Wainscoat}, {Chambers}, {Keane}, {Petric}, {Denneau},
  {Magnier}, {Berger}, {Huber}, {Flewelling}, {Waters}, {Schunova-Lilly}, \&
  {Chastel}}]{meewermic17}
{Meech}, K.~J., {Weryk}, R., {Micheli}, M., {et~al.} 2017, \nat, 552, 378

\bibitem[{{Menten} {et~al.}(2007){Menten}, {Reid}, {Forbrich}, \&
  {Brunthaler}}]{menreifor07}
{Menten}, K.~M., {Reid}, M.~J., {Forbrich}, J., \& {Brunthaler}, A. 2007, A\&A,
  474, 515

\bibitem[{{Micheli} {et~al.}(2018){Micheli}, {Farnocchia}, {Meech}, {Buie},
  {Hainaut}, {Prialnik}, {Sch{\"o}rghofer}, {Weaver}, {Chodas}, {Kleyna},
  {Weryk}, {Wainscoat}, {Ebeling}, {Keane}, {Chambers}, {Koschny}, \&
  {Petropoulos}}]{micfarmee18}
{Micheli}, M., {Farnocchia}, D., {Meech}, K.~J., {et~al.} 2018, \nat, 559, 223

\bibitem[{Mihalas \& Binney(1981)}]{mihbin81}
Mihalas, D., \& Binney, J. 1981, Galactic Astronomy (New York: W. H. Freeman
  and Co.)

\bibitem[{{Miyamoto} \& {Nagai}(1975)}]{miynag75}
{Miyamoto}, M., \& {Nagai}, R. 1975, Publ. Astr. Soc. Japan, 27, 533

\bibitem[{{Moro-Mart{\'{\i}}n} {et~al.}(2009){Moro-Mart{\'{\i}}n}, {Turner}, \&
  {Loeb}}]{morturloe09}
{Moro-Mart{\'{\i}}n}, A., {Turner}, E., \& {Loeb}, A. 2009, AJ, 704, 733

\bibitem[{{Mulders} {et~al.}(2017){Mulders}, {Pascucci}, {Manara}, {Testi},
  {Herczeg}, {Henning}, {Mohanty}, \& {Lodato}}]{mulpasman17}
{Mulders}, G.~D., {Pascucci}, I., {Manara}, C.~F., {et~al.} 2017, \apj, 847, 31

\bibitem[{{Murray} {et~al.}(2004){Murray}, {Weingartner}, \&
  {Capobianco}}]{murweicap04}
{Murray}, N., {Weingartner}, J.~C., \& {Capobianco}, C. 2004, \apj, 600, 804

\bibitem[{{Nguyen} {et~al.}(2012){Nguyen}, {Brandeker}, {van Kerkwijk}, \&
  {Jayawardhana}}]{ngubraker12}
{Nguyen}, D.~C., {Brandeker}, A., {van Kerkwijk}, M.~H., \& {Jayawardhana}, R.
  2012, ApJ, 745, 119

\bibitem[{{Oswalt} {et~al.}(2016){Oswalt}, {Sion}, \& {McCook}}]{oswsiomcc}
{Oswalt}, J.~B.~H.~T.~D., {Sion}, E.~M., \& {McCook}, G.~P. 2016, arXiv
  e-prints, arXiv:1606.01236

\bibitem[{{Palla} \& {Stahler}(2002)}]{palsta02}
{Palla}, F., \& {Stahler}, S.~W. 2002, \apj, 581, 1194

\bibitem[{{Pecaut} {et~al.}(2012){Pecaut}, {Mamajek}, \& {Bubar}}]{pecmambub12}
{Pecaut}, M.~J., {Mamajek}, E.~E., \& {Bubar}, E.~J. 2012, \apj, 746, 154

\bibitem[{{Portail} {et~al.}(2015){Portail}, {Wegg}, {Gerhard}, \&
  {Martinez-Valpuesta}}]{porwegger15}
{Portail}, M., {Wegg}, C., {Gerhard}, O., \& {Martinez-Valpuesta}, I. 2015,
  MNRAS, 448, 713

\bibitem[{{Portegies Zwart} {et~al.}(2018){Portegies Zwart}, {Torres},
  {Pelupessy}, {B{\'e}dorf}, \& {Cai}}]{portorpel18}
{Portegies Zwart}, S., {Torres}, S., {Pelupessy}, I., {B{\'e}dorf}, J., \&
  {Cai}, M.~X. 2018, \mnras, 479, L17

\bibitem[{{Preibisch} \& {Mamajek}(2008)}]{premam08}
{Preibisch}, T., \& {Mamajek}, E. 2008, {The Nearest OB Association:
  Scorpius-Centaurus (Sco OB2)}, ed. B.~{Reipurth}, 235

\bibitem[{{Rafikov}(2018)}]{raf18}
{Rafikov}, R. 2018, AJ, 861, 35

\bibitem[{{Reid} \& {Brunthaler}(2004)}]{reibru04}
{Reid}, M.~J., \& {Brunthaler}, A. 2004, ApJ, 616, 872

\bibitem[{{Reipurth}(2008)}]{rei08}
{Reipurth}, B. 2008, {Handbook of Star Forming Regions, Volume II: The Southern
  Sky}

\bibitem[{{Riedel} {et~al.}(2017){Riedel}, {Blunt}, {Lambrides}, {Rice},
  {Cruz}, \& {Faherty}}]{rieblulam18}
{Riedel}, A.~R., {Blunt}, S.~C., {Lambrides}, E.~L., {et~al.} 2017, \aj, 153,
  95

\bibitem[{{Riedel} {et~al.}(2018){Riedel}, {Blunt}, {Lambrides}, {Rice},
  {Cruz}, \& {Faherty}}]{rieblulamcatalog18}
---. 2018, VizieR Online Data Catalog, 515

\bibitem[{{Robin} {et~al.}(2017){Robin}, {Bienaym{\'e}},
  {Fern{\'a}ndez-Trincado}, \& {Reyl{\'e}}}]{robbiefer17}
{Robin}, A.~C., {Bienaym{\'e}}, O., {Fern{\'a}ndez-Trincado}, J.~G., \&
  {Reyl{\'e}}, C. 2017, \aap, 605, A1

\bibitem[{{Sch{\"o}nrich} {et~al.}(2010){Sch{\"o}nrich}, {Binney}, \&
  {Dehnen}}]{schbindeh10}
{Sch{\"o}nrich}, R., {Binney}, J., \& {Dehnen}, W. 2010, MNRAS, 403, 1829

\bibitem[{{Song} {et~al.}(2012){Song}, {Zuckerman}, \& {Bessell}}]{sonzucbes12}
{Song}, I., {Zuckerman}, B., \& {Bessell}, M.~S. 2012, \aj, 144, 8

\bibitem[{{Spezzi} {et~al.}(2008){Spezzi}, M., E., A., D., I., N., J., L., K.,
  B., \& R.}]{spealccov08}
{Spezzi}, L., M., A.~J., E., C., {et~al.} 2008, ApJ, 680, 1295.
\newblock \url{https://doi.org/10.1086%2F587931}

\bibitem[{{Stauffer} {et~al.}(2010){Stauffer}, {Tanner}, {Bryden}, {Ramirez},
  {Berriman}, {Ciardi}, {Kane}, {Mizusawa}, {Payne}, {Plavchan}, {von Braun},
  {Wyatt}, \& {Kirkpatrick}}]{statanbry10}
{Stauffer}, J., {Tanner}, A.~M., {Bryden}, G., {et~al.} 2010, \pasp, 122, 885

\bibitem[{{Szegedi-Elek} {et~al.}(2013){Szegedi-Elek}, {Kun}, {Reipurth},
  {P{\'a}l}, {Bal{\'a}zs}, \& {Willman}}]{szekunrei13}
{Szegedi-Elek}, E., {Kun}, M., {Reipurth}, B., {et~al.} 2013, Astrophys. J.
  Suppl., 208, 28

\bibitem[{{Torres} {et~al.}(2012){Torres}, {Loinard}, {Mioduszewski}, {Boden},
  {Franco-Hern{\'a}ndez}, {Vlemmings}, \& {Rodr{\'{\i}}guez}}]{torloimio12}
{Torres}, R.~M., {Loinard}, L., {Mioduszewski}, A.~J., {et~al.} 2012, ApJ, 747,
  18

\bibitem[{{Torres} {et~al.}(2007){Torres}, {Loinard}, {Mioduszewski}, \&
  {Rodr{\'{\i}}guez}}]{torloimio07}
{Torres}, R.~M., {Loinard}, L., {Mioduszewski}, A.~J., \& {Rodr{\'{\i}}guez},
  L.~F. 2007, ApJ, 671, 1813

\bibitem[{{Voirin} {et~al.}(2018){Voirin}, {Manara}, \& {Prusti}}]{voimanpru18}
{Voirin}, J., {Manara}, C.~F., \& {Prusti}, T. 2018, A\&A, 610, A64

\bibitem[{{Wenger} {et~al.}(2000){Wenger}, {Ochsenbein}, {Egret}, {Dubois},
  {Bonnarel}, {Borde}, {Genova}, {Jasniewicz}, {Lalo{\"e}}, {Lesteven}, \&
  {Monier}}]{wenochegr00}
{Wenger}, M., {Ochsenbein}, F., {Egret}, D., {et~al.} 2000, \aaps, 143, 9

\bibitem[{{Wielen}(1977)}]{wie77}
{Wielen}, R. 1977, \aap, 60, 263

\bibitem[{{Wright} \& {Mamajek}(2018)}]{wrimam18}
{Wright}, N.~J., \& {Mamajek}, E.~E. 2018, Monthly Notices of the Royal
  Astronomical Society, 476, 381.
\newblock \url{https://doi.org/10.1093/mnras/sty207}

\bibitem[{{Zhang}(2018)}]{zha18}
{Zhang}, Q. 2018, Astrophys. J. Lett., 852, L13

\bibitem[{{Zuluaga} {et~al.}(2018){Zuluaga}, {S{\'a}nchez-Hern{\'a}ndez},
  {Sucerquia}, \& {Ferr{\'\i}n}}]{zulsansuc18}
{Zuluaga}, J.~I., {S{\'a}nchez-Hern{\'a}ndez}, O., {Sucerquia}, M., \&
  {Ferr{\'\i}n}, I. 2018, \aj, 155, 236

\end{thebibliography}



\end{document}